\pdfoutput=1
\documentclass[twocolumn,secnumarabic,amssymb, nobibnotes, aps, pra, superscriptaddress, graphics, floatfix]{revtex4-1} 

\usepackage{graphicx}
\usepackage{amsmath,amssymb,bm}
\usepackage{upgreek}
\usepackage[driverfallback=dvipdfmx, colorlinks=true, linkcolor=blue, citecolor=blue, filecolor=blue, urlcolor=blue]{hyperref}

\begin{document}

\title{Attosecond Control of Electron Beams \\ at Dielectric and Absorbing Membranes}

\author{Yuya Morimoto}
\affiliation{Ludwig-Maximilians-Universit{\"a}t M{\"u}nchen, Am Coulombwall 1, 85748 Garching, Germany}
\affiliation{Max-Planck-Institute of Quantum Optics, Hans-Kopfermann-Str. 1, 85748 Garching, Germany}
\author{Peter Baum}
\email[]{peter.baum@lmu.de}
\affiliation{Ludwig-Maximilians-Universit{\"a}t M{\"u}nchen, Am Coulombwall 1, 85748 Garching, Germany}
\affiliation{Max-Planck-Institute of Quantum Optics, Hans-Kopfermann-Str. 1, 85748 Garching, Germany}

\date{January 23, 2018}

\begin{abstract}
Ultrashort electron pulses are crucial for time-resolved electron diffraction and microscopy of fundamental light-matter interaction.
In this work, we study experimentally and theoretically the generation and characterization of attosecond electron pulses by optical-field-driven compression and streaking at dielectric or absorbing interaction elements. 
The achievable acceleration and deflection gradient depends on the laser-electron angle, the laser's electric and magnetic field directions and the foil orientation. 
Electric and magnetic fields have similar contributions to the final effect and both need to be considered. 
Experiments and theory agree well and reveal the optimum conditions for highly efficient, velocity-matched electron-field interactions in longitudinal or transverse direction. 
We find that metallic membranes are optimum for light-electron control at mid-infrared or terahertz wavelengths, but dielectric membranes are excel in the visible/near-infrared regimes and are therefore ideal for the formation of attosecond electron pulses.
\end{abstract}


\maketitle



\section{Introduction}  \label{sec1}
Almost any light-matter interaction starts with the motion of charges in the oscillating electromagnetic cycles of light. A full visualization of such dynamics requires attosecond resolution in time and nanoscale/atomic resolution in space, which can be achieved by attosecond electron diffraction and microscopy\cite{Morimoto2017} with laser-generated, sub-relativistic attosecond electron pulse trains\cite{Morimoto2017, kozak2017, priebe2017}. More generally, the all-optical control of energy, momentum and spatiotemporal shape of free electrons with the cycles of laser light is of fundamental interest to quantum physics\cite{priebe2017, jones2016, andrick1976, weingartshofer1977, agostini1979, freimund2001,kruger2011, herink2012, feist2015}, electron pulse characterization\cite{hebeisen2006,kirchner2014}, ultrafast space-time imaging\cite{schoenlein2000,itatani2004,meckel2008,barwick2009,huismans2011,blaga2012,yurtsever2012,morimoto2014,morimoto2015,wolter2016,hassan2017}, quantum electron microscopy\cite{kruit2016}, free-electron lasers\cite{mcneil2010,andonian2011,hemsing2014,huang2016} and laser-based electron accelerators\cite{mizuno1987,peralta2013,breuer2013,nanni2015,carbajo2016}. Streaking of photoelectrons by laser cycles is also the basis of photon-based attosecond science with extreme-ultraviolet pulses\cite{paul2001,hentschel2001} and used, for example, for characterizing few-cycle laser pulses\cite{goulielmakis2004,kim2013,carpeggiani2017} and for investigating attosecond phenomena such as tunnelling, linear/nonlinear polarization and electron correlation with sub-cycle resolution in time\cite{cavalieri2007,eckle2008,sommer2016,ossiander2016}. 

Atomic-scale space-time imaging and diffraction with electrons require a de Broglie wavelength that is neither far longer nor far shorter than atomic distances. Convenient wavelengths are provided at sub-relativistic electron energies of, typically, 30-300 keV. In this regime, vacuum is dispersive, and creating ultrashort electron pulses implies the necessity for temporal compression even in the absence of space-charge effects\cite{aidelsburger2010}. Ideally, the compression of free electrons and their subsequent characterizations are all-optical\cite{kealhofer2016}, that is, based on laser-generated radiation, because it links the timing stability of pulsed lasers with the atomic-resolution capabilities of electrons\cite{baum2017ferenc}. However, the interaction of electrons with laser cycles is limited by the conservation of energy and momentum transfer\cite{edighoffer1979,park2010}. A free electron can only acquire energy/momentum directly from laser field cycles in the presence of a third body, for example atomic potentials\cite{andrick1976,weingartshofer1977,agostini1979,morimoto2014,morimoto2015,itatani2002,kitzler2002}, gratings\cite{kozak2017,mizuno1987,peralta2013,breuer2013}, nanoparticles\cite{barwick2009,yurtsever2012,hassan2017,piazza2015}, nano-tips\cite{feist2015} or metal foils\cite{kirchner2014,kealhofer2016,vanacore2017pre}. Only at very high optical intensities, where an electron can simultaneously absorb and emit multiple photons with different wave-vectors, there is ponderomotive scattering\cite{freimund2001,hebeisen2006,kozaknphys2017}. 

In this work, we consider the laser-optical control of free electrons with ultrathin dielectric and absorbing foils, that is, membranes with typically $\sim$100 nm thickness. A planar membrane is probably the simplest possible case of a third body, because it separates free space simply into two halves with otherwise unaltered characteristics. The first reports of such physics were made with metal membranes\cite{kirchner2014,plettner2005}. In contrast to nanostructures\cite{feist2015,barwick2009,yurtsever2012,piazza2015} or evanescent-wave geometries\cite{kozak2017,peralta2013,breuer2013}, a membrane does not require nanometer-thick electron beams but can be velocity-matched\cite{kirchner2014} for larger-diameter electron beams which are typical in table-top pump-probe experiments\cite{gerbig2015,waldecker2015,lahme2014} or at free-electron lasers. In contrast to metal membranes\cite{kirchner2014,kealhofer2016,vanacore2017pre,plettner2005} or the various nanostructures listed above, dielectrics are mostly non-absorbing and therefore have high laser-damage thresholds\cite{morimoto2017-damage}, enabling the application of highest-possible lasers intensities for electron control. The applicable electric peak fields that dielectrics can survive can be as high as the Coulombic fields between atoms ($\sim$1 V/$\rm{\AA}$) if the material is excited with few-cycle pulses\cite{schultze2014,luu2015}. Various types of dielectric foils can be manufactured as free-standing membranes with thicknesses below 100 nm, where sub-relativistic electrons can pass through. Dimensions of $>$100 $\upmu$m of lateral size are feasible\cite{morimoto2017-damage} and big enough to cover almost any available type of electron beam. 

This work reports a combined experimental and theoretical study of dielectric and absorbing membranes for laser-electron control. We identify the relevant physics for electron pulse compression, streaking deflection and acceleration in dependence on the foil material, membrane thickness, laser polarization, peak field amplitude and the angles involved. In Section \ref{sec2}, we discuss the basic mechanisms in dielectrics that mediate field-driven acceleration and deflection. We report an analytical model that predicts the outcome of time-dependent sub-cycle interaction for arbitrary interaction geometries. In Section \ref{sec3}, we compare our theory to a set of experimental results regarding sub-cycle deflection, finding remarkable agreement. We also report the simultaneous streaking of multiple Bragg reflections, further establishing the feasibility of attosecond-Angstrom diffraction beyond our initial report\cite{Morimoto2017}. In Section \ref{sec4}, we report experimental results on the temporal compression of free electron wavepackets to attosecond duration with various types of dielectric membranes. We discuss the fundamental and practical limits of this approach and report the necessary experimental procedures to reach a quantum-limited pulse duration in practice. In Section \ref{sec5}, we discuss the potential use of absorbing materials. In Section \ref{sec6}, we discuss the importance of magnetic fields. We conclude in Section \ref{sec7} by summarizing our results and also give an outlook on future perspectives we see emerging from the reported results on light-electron control. 

\section{Theory \label{sec2}}
\begin{figure}[!tb]
  \includegraphics[width=8.5cm]{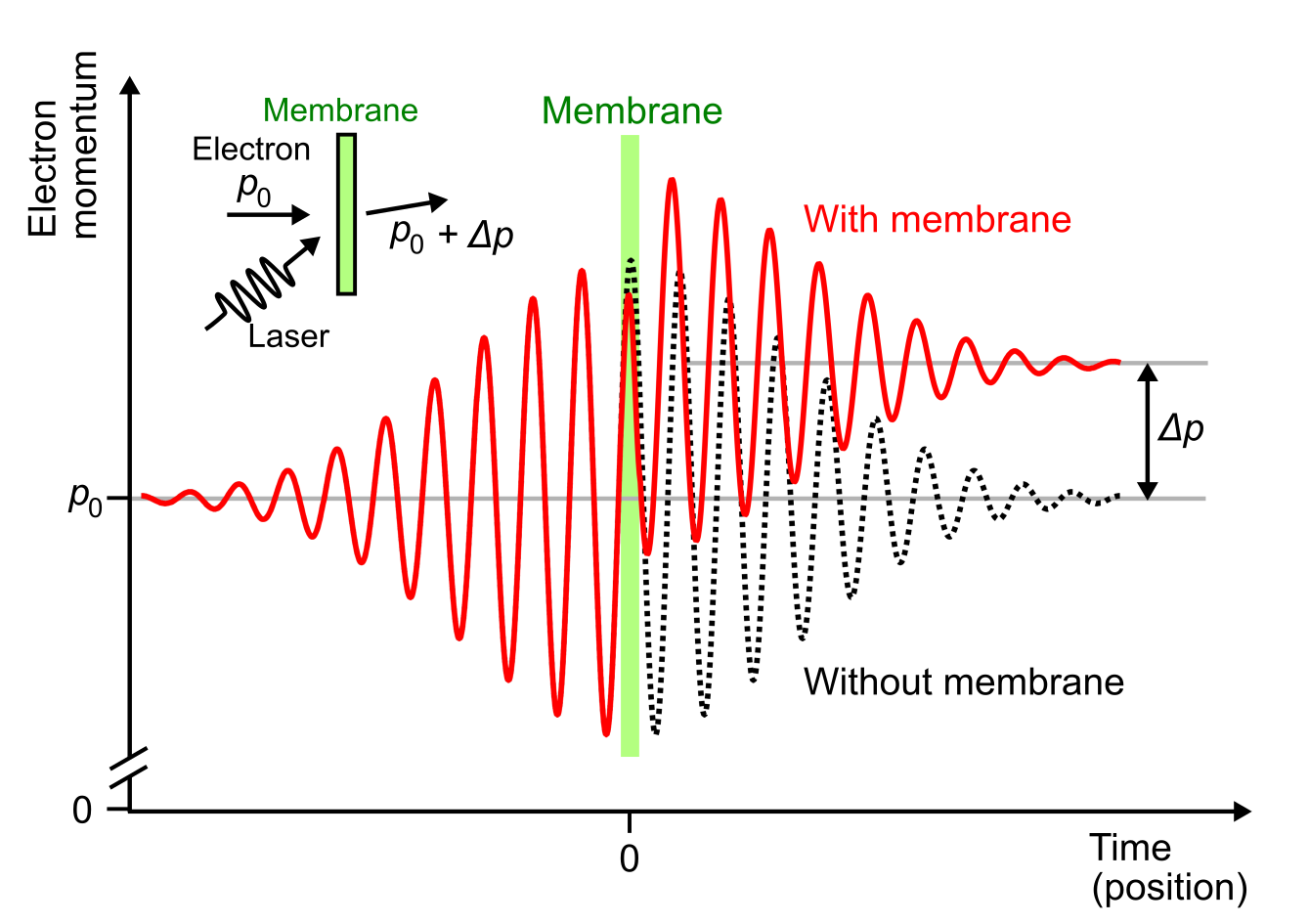}%
  \caption{Concept of attosecond control of free electrons at dielectric or absorbing membranes. Without a membrane (green), the electron momentum (black dotted curve) returns to the initial value ($p_0$) after interaction with a laser field. In contrast, the electron acquires a finite momentum $\Delta p$ (red curve) in the presence of a membrane, mainly due to a phase shift of the oscillations caused by the refractive index and thin-film interferences. The actual momentum shift is a vectorial quantity that depends on various angles and polarizations, as investigated in this work. \label{fig1}}
\end{figure}

\begin{figure}[!tb]
  \includegraphics[width=8.2cm]{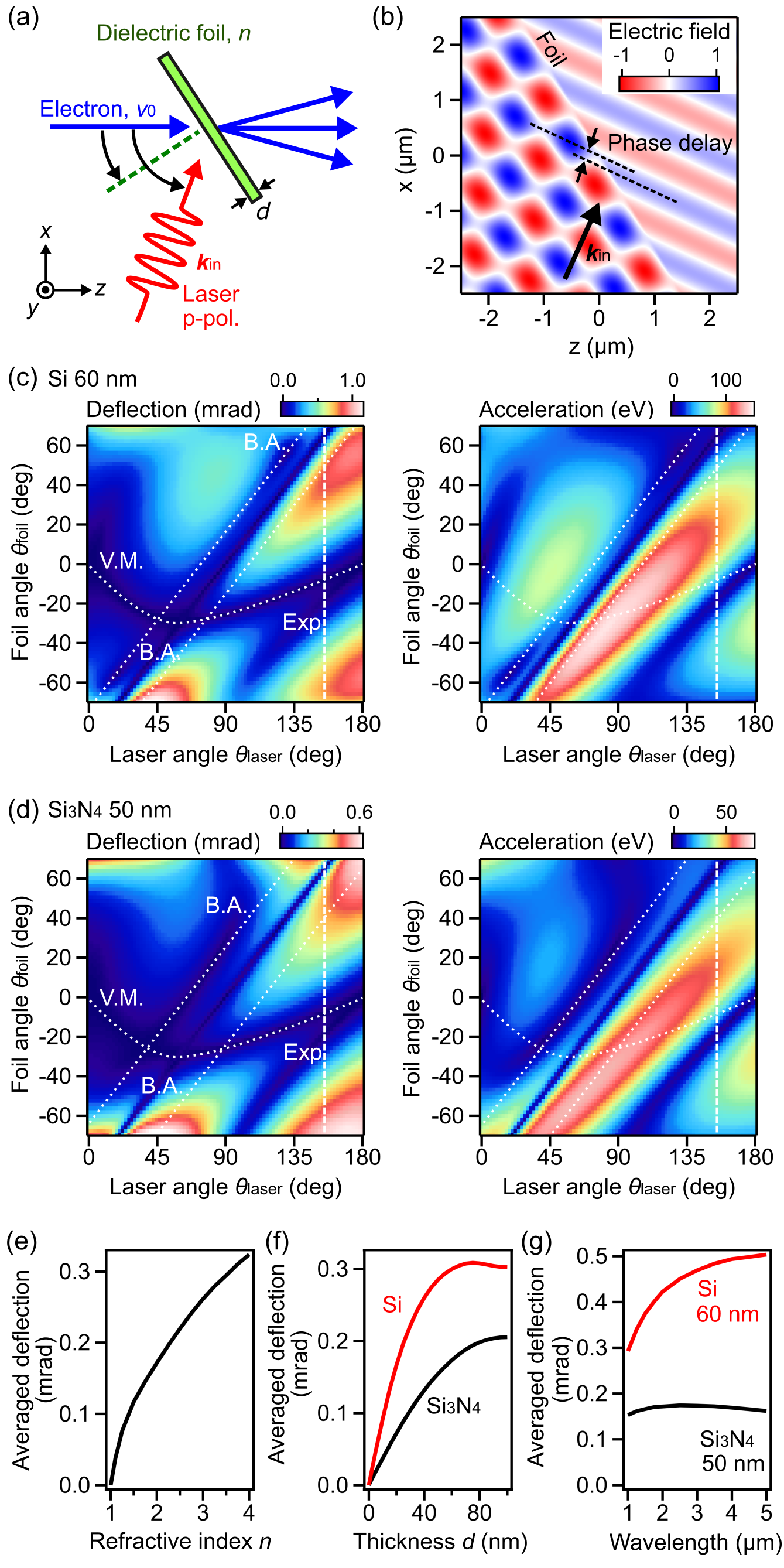}%
  \caption{Theory for laser streaking and acceleration/deceleration of free electrons with dielectric foils. (a) Angle definitions. (b) Time-frozen snapshot of a p-polarized electric field around a 60-nm-thin Si membrane. There is a phase shift (dotted lines) between the crests in front and behind the foil. (c) Peak deflection and peak acceleration for a 60-nm-thick Si foil. (d) Peak deflection and peak acceleration for a 50-nm-thick Si$_3$N$_4$ foil. The incident laser field has a wavelength of 1030 nm and a peak field strength of 1 GV/m and 3.3 T. White dotted lines represent angle combinations that provide velocity matching (V.M.) or Brewster's angle (B.A.). The white dashed line (Exp) denotes the conditions used in most of the experiments. (e) Mean deflection as a function of refractive index. (f) Mean deflection as a function of membrane thickness. (f) Mean deflection as a function of laser wavelength. \label{fig2}}
\end{figure}

In this section, we introduce a classical model which describes the propagation of a free electron passing through a membrane in the presence of a laser field. 
The inset of Fig. \ref{fig1} and Fig. \ref{fig2}(a) depict our theoretical approach. The electrons are point particles and the laser field is a pulsed electromagnetic plane-wave that is modified by the reflection, transmission, refraction, absorption and the interferences around and inside of the foil, assumed to be planar and of constant thickness and index of refraction. 
Electrons are assumed to travel at a constant velocity $v_0$ (typically 0.1-0.9 times the speed of light) along a linear trajectory from $z = -\infty$ to $z = +\infty$, corresponding to initial and final times at which the pulsed laser field is zero. The laser pulse duration is assumed to be sufficiently longer than an optical-cycle, which allows us to obtain an analytical form of the electromagnetic fields around and inside the membrane (see Appendix \ref{seca1}). 

Figure \ref{fig1} explains the basic physics of the interaction in a simplified depiction. Without a foil, in free space, there would be periodic momentum oscillations driven by the electric and magnetic field cycles, but these would cancel out after the pulse has passed in time (see Fig. \ref{fig1}, dotted line). In other words, the electron does not obtain any net momentum from the field cycles; left are only the much weaker ponderomotive effects that are not subject of this paper. In contrast, if the laser field impinges on a dielectric or absorbing foil, there occur refraction, thin-film interferences, partial reflection, absorption and transmission. Also, the optical wavelength is reduced by the refractive index inside of the foil. The electromagnetic fields on each side can therefore acquire different strengths and phases. For the electron that passes through the membrane in sub-femtosecond time, these effects break the periodicity of the cycle-driven momentum oscillations and therefore cause an overall net momentum change $\Delta p$ after the interaction is over (see Fig. \ref{fig1}, red line), although there is neither necessarily an abrupt field cancelation like at metal mirrors\cite{kirchner2014,plettner2005} or an optical near-field like at nanostructures\cite{barwick2009}.

\subsection{Formalism \label{sec2a}}
Figure \ref{fig2}(a) depicts the geometry and angle definitions that are used in the following discussion. For evaluation of the overall momentum gain $\Delta \boldsymbol{p}$ as a function of the membrane geometry and way of optical excitation, we integrate for a propagating electron the time-dependent Lorentz force over time: 
\begin{eqnarray}
\Delta \boldsymbol{p}(t_0) =&& -e\int_{-\infty}^{\infty} \boldsymbol{E}( \boldsymbol{r}_{\rm{e}}(t-t_0),t) {\rm{d}}t  \nonumber \\
&& -e \int_{-\infty}^{\infty} \boldsymbol{v}_{\rm{e}} \times \boldsymbol{B}( \boldsymbol{r}_{\rm{e}}(t-t_0),t) {\rm{d}}t.
\label{eq1}
\end{eqnarray}
Here, $e$ is elementary charge, $t$ is time, $\boldsymbol{r}_{\rm{e}}$ and $\boldsymbol{v}_{\rm{e}}$  are the position and the velocity of the electron. The parameter $t_0$ is the time at which the electron reaches the middle of the foil ($z = 0$) and serves for studying time-dependent effects, for example the overall deflections or accelerations of electrons that hit the foil at different times. The vectors $\boldsymbol{E}(\boldsymbol{r}_{\rm{e}},t)$ and $\boldsymbol{B}(\boldsymbol{r}_{\rm{e}},t)$ are the electric and magnetic fields at $\boldsymbol{r}_{\rm{e}}$, respectively, and obtained by analytical calculation of the plane-wave field interferences around and inside the foil (see Appendix \ref{seca1}). Figure \ref{fig2}(b) illustrates the electric part of such a field, frozen in time, for $\theta_{\rm{foil}} = 30^\circ$ and $\theta_{\rm{foil}} = 65^\circ$. The magnetic part is proportional with exception of inside the foil. The dotted line depicts the effective phase shift between the lines of constant peak field on the two different sides, caused by the foil's refractive index and thin film interferences. We assume that the electrons are monochromatic with an initial electron velocity $v_0$ and travel exactly along the $z$-axis. We also assume that any light-induced change of momentum is much smaller than the initial electron momentum, that is, the electrons do not wiggle in space by more than a negligible fraction of the optical wavelength. In this picture, we can write $\boldsymbol{v}_{\rm{e}}=(0,0,v_0)$ and $\boldsymbol{r}_{\rm{e}} (t)=(0,0,v_0 t)$ and only need to consider the electromagnetic fields on the $z$-axis, denoted in the following as $\boldsymbol{E}(z,t)$ and $\boldsymbol{B}(z,t)$. 

It is useful to separate the final momentum change vector $\Delta \boldsymbol{p}$ into forward and sideways components. The forward momentum change $\Delta p_z$ implies an overall velocity change that, if dependent on the electron's arrival time $t_0$, can cause the compression of an electron beam into ultrashort pulses after some dispersive propagation\cite{gliserin2015}. The sideways momentum changes $\Delta p_x$ and $\Delta p_y$ imply deflection of the beam into directions perpendicular to the propagation axis. If $\Delta p_x$ and $\Delta p_y$ depend on $t_0$, this implies a time-dependent streaking like in a cathode-ray oscilloscope at optical frequencies1 which is useful for characterizing femtosecond or attosecond electron pulses\cite{Morimoto2017,kealhofer2016} or allows sub-cycle imaging of electromagnetic waveforms\cite{Morimoto2017, Ryabov2016}. In Appendix \ref{seca1}, we show that $\Delta p_x$, $\Delta p_y$ and $\Delta p_z$ indeed depend on $t_0$ and are sinusoidal functions at the frequency of the laser field.

We first consider the longitudinal momentum change $\Delta p_z$ after light-electron interaction. Magnetic field components can never accelerate or decelerate an electron; in other words, the longitudinal projection (along $z$-axis) of the second term in Eq. \ref{eq1}, $\boldsymbol{v}_{\rm{e}}\times \boldsymbol{B}$, is always zero. Therefore the longitudinal momentum is only affected by the $z$-component of the electric field. We obtain
\begin{eqnarray}
&&\Delta p_z(t_0) = -e\int_{-\infty}^{\infty} E_z( v_0 t - v_0 t_0,t) {\rm{d}}t  \nonumber \\
&&= -e\int_{-\infty}^{-T_{\rm{f}}/2+t_0} E_{z,\rm{vac}} ( v_0 t - v_0 t_0,t) {\rm{d}}t \nonumber \\
&& \ -e\int_{-T_{\rm{f}}/2+t_0}^{T_{\rm{f}}/2+t_0} E_{z,\rm{foil}} ( v_0 t - v_0 t_0,t) {\rm{d}}t \nonumber \\
&& \ -e\int_{T_{\rm{f}}/2+t_0}^{\infty} E_{z,\rm{vac}} ( v_0 t - v_0 t_0,t) {\rm{d}}t. 
\label{eq2}
\end{eqnarray}
Here, we separated the integral of Eq. \ref{eq1} into three parts, one each for the propagation before, inside and behind the foil. $T_{\rm{f}}$ is the duration that the electron spends inside of the foil, typically attoseconds for nanometer-thin membranes. $E_{z,\rm{vac}} (z,t)$ and $E_{z,\rm{foil}} (z,t)$ are $z$-components of the laser electric fields in vacuum and inside the foil, respectively. It is also instructive and useful in practice to convert $\Delta p_z$ into the energy domain. If the momentum change $\Delta p_z$ is much smaller than the initial momentum $p_0$, and if sideways deflections are negligible, the energy change $\Delta W$ is proportional to $\Delta p_z$, because $\Delta W \approx p_0 c^2 \Delta p_z/\sqrt{p_0^2 c^2+m_{\rm{e}}^2 c^4}$, where $c$ is the speed of light and $m_{\rm{e}}$ is the rest mass of the electrons.

Second, we consider the sideways momentum changes $\Delta p_x$ and $\Delta p_y$ which cause a deflection perpendicular to the beam's propagation axis after light-electron interaction. In a similar way as above, the deflection is determined by time integrals of the electric and magnetic fields. We obtain 
\begin{eqnarray}
\Delta p_x (t_0) =&& -e \int_{-\infty}^{\infty} E_x ( v_0 t - v_0 t_0,t) {\rm{d}}t \nonumber \\
&& +e v_0 \int_{-\infty}^{\infty} B_y (v_0t-v_0t_0,t) {\rm{d}}t,
\label{eq3}
\end{eqnarray}
\begin{eqnarray}
\Delta p_y (t_0) =&& -e\int_{-\infty}^{\infty} E_y( v_0 t - v_0 t_0,t) {\rm{d}}t \nonumber \\
&& -e v_0 \int_{-\infty}^{\infty} B_x (v_0t-v_0t_0,t) {\rm{d}}t.
\label{eq4}
\end{eqnarray}
Like in Eq. \ref{eq2}, each integral can be decomposed into three terms, namely propagation before the foil, inside, and afterwards. The electromagnetic fields and their integrals in Eqs. \ref{eq2}-\ref{eq4} are evaluated by considering infinite series of monochromatic plane waves; see the Appendix \ref{seca1} for details.

\subsection{Consequences of optical polarization \label{sec2b}}
Before reporting detailed results, we quickly comment on polarization. For p-polarized incident light ($E_y=B_x=B_z=0$) deflection along the $y$-axis is zero. 
For s-polarized light ($E_x=E_z=B_y=0$), there is never any acceleration/compression, because there are no electric field components along the electron propagation direction. There is also no deflection along the $x$-axis. Overall, there can only be acceleration/compression or deflection/streaking along a direction that projects at least in part onto the direction of the electric field vector. 

Interestingly, as discussed below by theory and experiment, we find for the case of s-polarization that deflection along the$y$-axis is also very close to zero in all cases. This result, as we will see below, is caused by an almost perfect cancellation of the contributions from electric and magnetic fields after the interaction, although the electron momentum does indeed oscillate considerably while being in the laser fields. As a result, we find that s-polarized light does not change the electron momentum along any direction regardless of the membrane type and incidence angles. The use of p-polarization is the only practical way for achieving electron pulse compression in time or for pulse characterization by streaking metrology.

\subsection{Phenomenological description for dielectrics; phase changes and inner-foil effects \label{sec2c}}
For the case of non-absorbing dielectrics, which are most useful for almost all kinds of experiments (see below), it may be instructive to separate the physics into two phenomenological contributions.
Inside of the foil, the optical electric field strength is reduced via the refractive index; only the magnetic fields retain their full strength, because almost any dielectric material has a relative magnetic permeability of $\mu/\mu_0 \approx 1$ at optical frequencies. The passing electron experiences for a short time (passing through the foil) a reduction of momentum change as compared to the free-space case. Outside of the foil, there are fields on both sides, but due to the foil's refractive index and the thin-film optical interferences, there is in almost all cases an optical phase shift between the waves on both sides. An electron passing through the foil is therefore injected into a field of different temporal phase on the other side and integration results in a non-zero momentum shift. For example, at Brewster's angle, the overall momentum change is non-zero, although there are no reflections or absorptions of any kind, and intensity on both sides is equal. This out-of-the-foil, phase-change mechanism is typically much stronger as the inside-of-the-foil effect, as we will see. Magnetic and electric fields contribute almost equally to the maximum amplitude modulation for sub-relativistic electrons, but with different dependencies on angles and polarizations. 

\subsection{Theoretical results on dielectric membranes \label{sec2d}}
First, we concentrate on non-absorbing dielectric materials; absorbers will be discussed in Section \ref{sec5}. The above reported formalism allows to use a simple analytical expression for the electromagnetic fields of a plane wave (see Appendix \ref{seca1}) to predict the deflection and acceleration of an electron beam in dependence on the interaction geometry. Figure \ref{fig2} reports some results. The kinetic energy of the incident electrons is 70 keV. We consider a p-polarized incident laser field at a wavelength of 1030 nm and a peak field strength of 1 GV/m and 3.3 T in vacuum. Figures \ref{fig2}(c) and \ref{fig2}(d) show the theoretical results for two cases, first, for a 60-nm-thick Si membrane (refractive index $n = 3.6$) and second, for a 50-nm-thick $\rm{Si}_3\rm{N}_4$ foil ($n = 2.0$); see Fig. \ref{fig2}(a) for the angle definitions. We plotted in these figures the peak deflection or peak acceleration that is achieved in a long electron pulse, that is, the maximum while varying $t_0$.

The first observation is that deflection and acceleration/compression occur simultaneously at many conditions. This finding has substantial consequences for practical pulse compression; see Section \ref{sec4} and is also related to some resolution degradations in near-field electron microscopy of nanostructures\cite{plemmons2016}. However, as we will see, there are a lot of favorable conditions to choose from in order to avoid any undesired complexity; see Sections \ref{sec3}, \ref{sec6} and below.

The white lines marked as B.A. in Figs. \ref{fig2}(c) and \ref{fig2}(d) denote Brewster's angle, where there are no optical reflections at both membrane surfaces for the 1030-nm light. This case is interesting, because it can be excluded that reflections or instantaneous changes of intensity, as previously observed in metal foils\cite{kirchner2014,kealhofer2016}, could explain the physics of deflection for dielectric membranes. Indeed, although all light is entirely transmitted, there is a substantial deflection ($\Delta p_x$) and acceleration ($\Delta p_z$) induced even at Brewster's angles, showing the importance of the optical phase delay as explained above. The dark diagonal line in the data at $\theta_{\rm{foil}} \approx \theta_{\rm{laser}} -90^\circ$ represents grazing incidence, that is, the angle at which the laser direction $\boldsymbol{k}_{\rm{in}}$ and the membrane surface are parallel. This case is impractical, and in addition, there is almost no effect. Interestingly, the areas of highest effect strengths are found for both contributions, deflection and acceleration, at the lower-right region ($\theta_{\rm{foil}} < \theta_{\rm{laser}} -90^\circ$), where electrons and laser come from different sides of the foil, as compared to the upper-left part ($\theta_{\rm{foil}} > \theta_{\rm{laser}} -90^\circ$), where electrons and laser hit the same side of the foil. In a counter-propagating geometry, the optical phase delay generated by the refractive index and thin-film interferences adds to the time delay of the electron passing through the foil, while in co-propagating geometry the two effects subtract from each other (see Appendix \ref{seca2} for details). 

Intriguingly, at angles satisfying velocity-matching condition (white curves marked as V.M.), where the relative timing between electrons and field-cycles is independent on the beam diameter ($x$-coordinate)\cite{kirchner2014,williamson1993,baum2006}, our theory predicts almost zero sideways deflection due to cancellation of electric and magnetic components (see Section \ref{sec6}), although a lot of different forces are contributing to the overall interaction. This finding, which has been theoretically reported before for streaking by metal foils\cite{andrey_thesis2017}, is of great importance to experiments, because it implies that all-optical compression of large-diameter electron pulses with THz radiation\cite{kealhofer2016} or optical field cycles\cite{Morimoto2017,priebe2017} can indeed produce perfectly non-tilted and non-divergent pulses if velocity matching angles are applied. Attosecond diffraction and microscopy applications\cite{Morimoto2017} are therefore feasible with large-diameter beams in table-top experiments. On the down side, streaking metrology of non-tilted, large beams is difficult without a resonator element\cite{kealhofer2016} or array thereof. 

This overall physical picture does not strongly depend on the dielectric membrane's material or foil thickness. When we relate Figs. \ref{fig2}(c) and \ref{fig2}(d), which are the results for Si and $\rm{Si}_3\rm{N}_4$, respectively, we find that the overall features are rather similar, although the peak deflection and acceleration is different by a factor of $\sim$2, caused by the differences in refractive index and foil thickness. In order to clarify these influences on the overall effect, we plot in Figs. \ref{fig2}(e) and \ref{fig2}(f) the sideways deflection as a function of a foil's refractive index $n$ and foil thickness $d$, respectively. Here, we averaged over all angle combinations plotted in the above two-dimensional figures ($\theta_{\rm{laser}} = 0^\circ \dots  180^\circ$, $\theta_{\rm{foil}} = -70^\circ \dots 70^\circ$). In Fig. \ref{fig2}(f), we assumed that $n$ is constant for all the thickness. We find that a higher refractive index gives stronger deflection because it causes a larger phase delay. Likewise, thicker membranes (up to a certain limit) produce stronger deflection via the longer inside paths for both laser and electrons, producing also a larger phase delay. The onset of saturations can be seen at large refractive indices above $\sim$5 and for foil thicknesses approaching a substantial fraction of the wavelength inside of the material, especially for the Si foil. This causes the initial regions of strongest deflection or acceleration to diminish or even to become zero. This cancelation at some angles within the averaged ranges leads to partial reduction of the overall effect. 

Figure \ref{fig2}(g) shows the laser-wavelength dependence for otherwise constant parameters, again plotted as the average value for all angle combinations. At constant peak field strength, the time-integral of the Lorentz force and therefore the peak oscillation amplitude of the electron momentum increase proportionally to wavelength. On the other hand, the optical phase delay caused by the foil usually decreases with wavelength, if assuming that the refractive index stays approximately constant. These two effects cancel out each other to some extent. For example, the deflection by the $\rm{Si}_3\rm{N}_4$ foil (black line) is nearly independent on laser wavelength; see Fig. \ref{fig2}(g). This is in noticeable contrast to streaking by metallic foils, in which the deflection is directly proportional to wavelength\cite{kirchner2014,andrey_thesis2017} (see also Section \ref{sec5}). This difference is due to the dissimilar physics being responsible, namely abrupt field cancelation for metals vs. phase change effects for dielectrics. The deflection by the Si foil (red line) is also somewhat constant at longer wavelengths. 

In summary, the here reported simple model described by Eqs. \ref{eq2}-\ref{eq4} predicts a rich set of angular combinations for practical purposes for electron pulse compression or streaking deflection. There are also two remarkable findings, namely no sideways deflection at velocity-matching, and a substantially stronger deflection and compression for laser/electron incidences from different directions.

\subsection{Electromagnetic fields inside dielectric membranes \label{sec2e}}

\begin{figure}[bt]
  \includegraphics[width=8.5cm]{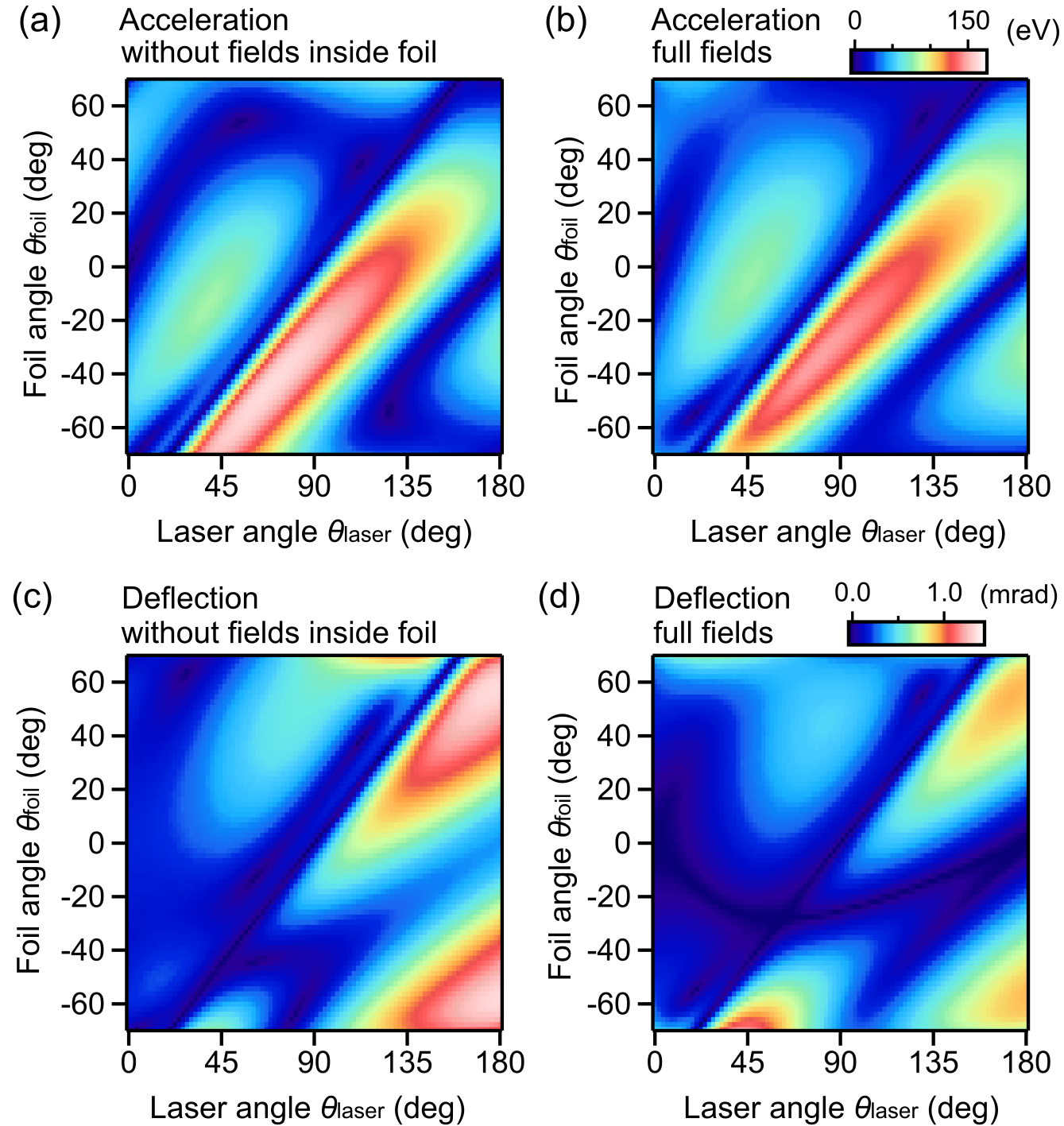}%
  \caption{Contribution of the effects outside and inside of the membrane. (a) Acceleration approximated by assuming no effect inside the foil. (b) Acceleration with full calculation. (c) Deflection with no effect inside the foil. (d) Deflection with full calculation. We assume a p-polarized incident field of 1 GV/m at 1030-nm wavelength and a 60-nm-thin Si membrane. There is almost no difference; see Section \ref{sec2d}. \label{fig3}}
\end{figure}

In Section \ref{sec2c}, we had argued that in many cases the overall effect in dielectric membranes can be described by a simple effective phase change of the incoming and outgoing waves while neglecting the dynamics inside the foil. In Eq. \ref{eq2}, this approximation corresponds to neglecting the second term in the second line. In order to confirm this prediction, we show in Figs. \ref{fig3}(a) and \ref{fig3}(c) the peak acceleration and deflection without the fields inside membranes, in comparison to Figs. \ref{fig3}(b) and \ref{fig3}(d), where the entire propagation is modeled completely. The difference is $<$20 $\%$ in acceleration and $<$50 $\%$ in deflection for most angle combinations. This result supports the simple picture of Section \ref{sec2c}, but compared to the acceleration, the results of deflection show somewhat larger deviations. While longitudinal momentum changes (acceleration/deceleration) can only originate from electric fields, deflection is caused by electric and magnetic fields in unison. Unlike the electric fields, magnetic fields are not reduced by the refractive index inside of the membrane. Deflection therefore depends for many angle combinations in a stronger way on the inner-foil dynamics than acceleration, for which a simple phase-change mechanism is a valid approximation (see Section \ref{sec2c}). 

\section{EXPERIMENT 1: SUB-CYCLE DEFLECTION \label{sec3}}
In this section, we report a set of experimental results on sub-optical-cycle sideways deflection for different dielectric foils and for different geometries. We investigate the deflection amplitudes as a function of incoming field strengths, laser polarization and foil angles. All the experimental results are well reproduced by the theory introduced in the previous section and therefore confirm the above made interpretations.

\subsection{Experimental setup and samples \label{sec3a}}
\begin{figure*}[tb]
  \includegraphics[width=17cm]{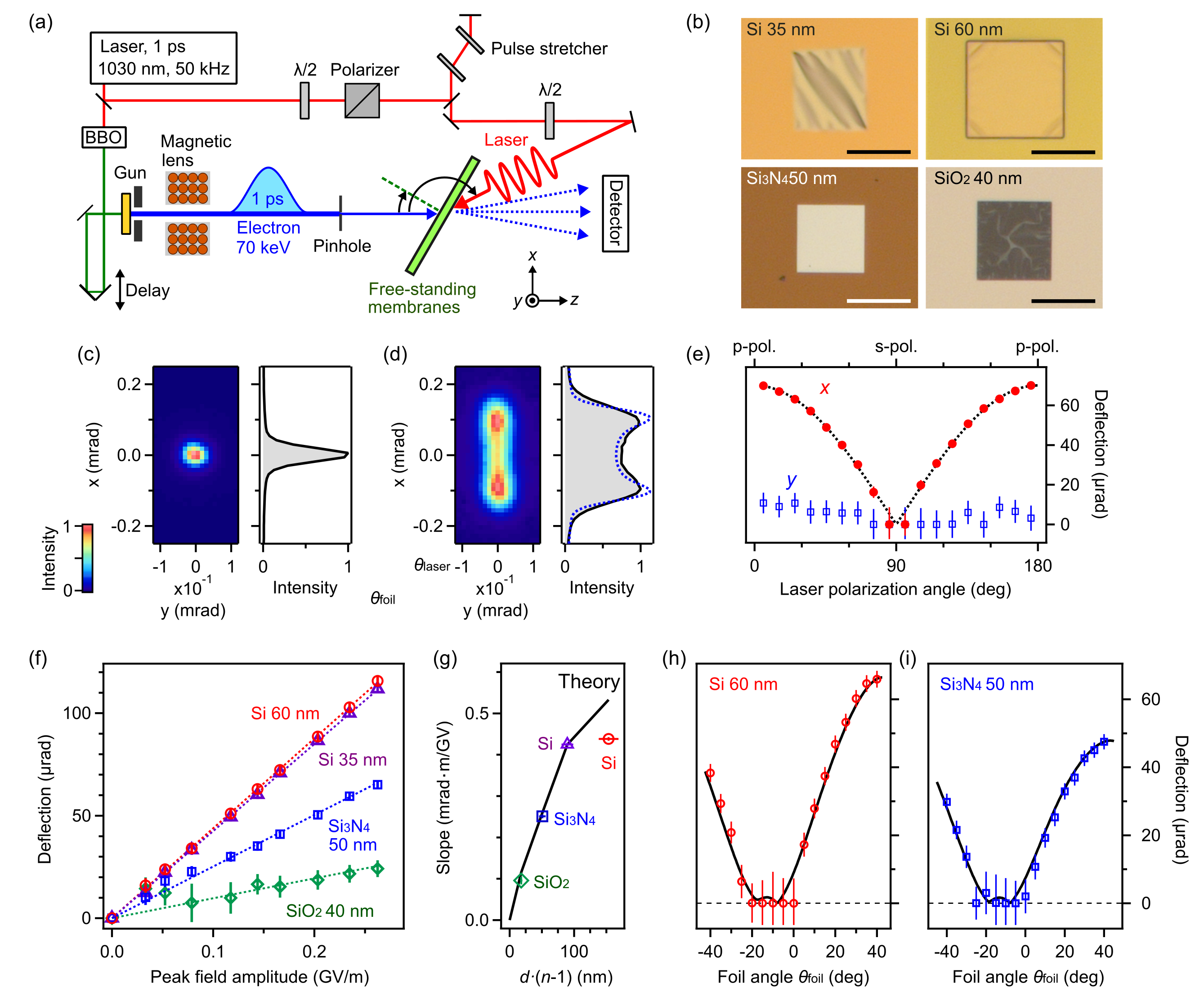}%
  \caption{Experimental results on attosecond sideways deflection. (a) Experimental set-up. $\lambda/2$, half-wave plate; BBO, $\beta$-barium-borate crystal for second-harmonic generation; yellow, photocathode. (b) Photographs of the free-standing membranes (dark-field microscopy). Scale bar, 100 $\upmu$m. (c) Electron beam profile. (d) Electron beam profile with a deflecting laser field. Blue dotted line, simulation. (e) Polarization-dependence of the deflection along $x$ (red dots) and $y$ (blue squares) in comparison to theory (dotted black line). (f) Measured deflection (symbols) as a function of the incident peak field strength in vacuum for different foil materials in comparison to theory (dotted lines). (g) Dependence of deflection rate on foil thickness $d$ and refractive index $n$. (h) Measured deflection amplitude (red circles) for a Si membrane as a function of the foil angle in comparison to theory (solid line). (h) Measured deflection amplitude (blue squares) for a 50-nm-thick $\rm{Si}_3\rm{N}_4$ membrane in comparison to theory (solid line).  \label{fig4}}
\end{figure*}

Figure \ref{fig4}(a) shows the experimental set-up. A regenerative Yb:YAG disk amplifier\cite{schneider2014} generated 1-ps pulses at 1030-nm wavelength with a repetition rate of 50 kHz. A small part of the laser output beam (red) was converted to its second-harmonic wavelength (green) and used for the ultrafast photoemission of electrons from a 20-nm-thick gold photocathode\cite{kasmi2015} (yellow). The photoelectrons were accelerated by an electrostatic field of $\sim$2 kV/mm to a final energy of 70 kV. This electron energy corresponds to a speed of 0.48 times the speed of light and a de Broglie wavelength of 4.5 pm. A magnetic lens (orange) approximately collimated the electron beam (blue) and guided it through the sample foils (green) onto a single-electron detector\cite{kealhofer2015a} (TVIPS GmbH) that was located at $\sim$2 m distance. The sample foils were placed at $\sim$0.5 m distance from the magnetic lens. If not otherwise specified, the different foils were all oriented at $\theta_{\rm{foil}} = 30^\circ$. To limit the size of the electron beam to the available foil dimensions (roughly 100$\times$100 $\upmu$m$^2$), we placed a pinhole (black) of 150 $\upmu$m diameter at $\sim$10 cm distance before the sample location. In order to avoid any space-charge effects, we used an average electron flux of $\sim$1 electron/pulse at the foils. The electron pulse duration was about 1 ps (full-width-at-half-maximum, FWHM), characterized by THz streaking\cite{kealhofer2016}. A second part of the laser output pulses at 1030 nm wavelength was used for the sideways deflection. The power was controlled by a half-wave plate ($\lambda$/2) and a polarizer. The laser pulses, stretched to 1.7 ps duration (FWHM) by a grating pair (1000 lines/mm, grey), were focused to $\sim$300 $\upmu$m diameter (1/$\rm{e}^2$ full width) onto the backside of the sample foils at an angle of $\theta_{\rm{laser}} = 155^\circ$. A half-wave plate ($\lambda$/2) was used to set the excitation polarization. The temporal stretching and large focus diameter of the optical pulses provided spatially and temporally homogeneous electromagnetic field cycles for all electrons in the beam. Images of the electron beam after the foil were taken with an exposure time of 1-2 s; the data shown below was averaged over tens of images. The relative delay time between electron and laser pulses was adjusted to maximize the observed sideways deflection. We measured a cross-correlation width of $>$2 ps here, as expected.

We applied four different free-standing dielectric membranes, namely 35 nm of Si, 60 nm of Si, 50 nm of $\rm{Si}_3\rm{N}_4$ and 40 nm of $\rm{SiO}_2$. The Si membranes were single-crystalline with $\langle100\rangle$ orientation. The $\rm{Si}_3\rm{N}_4$ and $\rm{SiO}_2$ foils were rather amorphous or slightly polycrystalline. Figure \ref{fig4}(b) shows dark-field microscope images of the membranes. The 60-nm-thick Si and $\rm{Si}_3\rm{N}_4$ foils were optically flat while the 35-nm-thick Si and $\rm{SiO}_2$ foils had some residual winkles. The 60-nm-thick Si foil was acquired from Norcada Inc. while the other foils were obtained from TEMwindows.com. The 60-nm-thick Si foil had a free-standing area of 150$\times$150 $\upmu$m$^2$, while the other foils had 100$\times$100 $\upmu$m$^2$ of free-standing region. At a foil angle of $\theta_{\rm {foil}} = 30^\circ$, the transmittance of our 70-keV electrons was $\sim$35$\%$, $\sim$30$\%$, $\sim$20$\%$ and $\sim$30$\%$ for 35-nm-thick Si, 60-nm-thick Si, 50-nm-$\rm{Si}_3\rm{N}_4$ and 40-nm-thick ${\rm SiO}_2$, respectively. There was only a minor fraction of inelastically scattered electrons visible on the screen as a faint halo round the central beam.

We found that the measured deflection amplitudes in all experiments were always $\sim$30$\%$ less than calculated from the optically determined peak field strengths, although all laser parameters (focal diameter, pulse duration, power) were characterized in the best possible way. This slight discrepancy, global to all results in this paper, is attributed to the non-Gaussian shapes of the beams, residual off-center misalignments or jitter of the experimental conditions. Also, although the laser pulses were made larger and longer than the electron pulses, there are residual distributions of field strengths over the electron beam profile. We therefore applied a constant factor to all figures where theory and experiment are compared; details are described below. 

\subsection{Experiments on laser-cycle deflection \label{sec3b}}
Figures \ref{fig4}(c) and \ref{fig4}(d) depict images of the electron beam after passing though the 60-nm Si foil without and with an incident p-polarized laser field ($2.6\times10^8$ V/m) in the $xz$ plane. We observe a pronounced elongation of the spot along the $x$-axis. This streaked image showed two spots with maximum intensity on both ends, which demonstrates that sideways deflection is indeed induced sinusoidally in time, as expected. The black curves in the right panels of Figs. \ref{fig4}(c) and \ref{fig4}(d) show the intensity profiles obtained by integrating the images over the $y$-direction. The blue, dotted curve in Fig. \ref{fig4}(d) is a result of a fitting via
\begin{eqnarray}
I_{\rm def} (x) = \frac{1}{T} \int_{-T/2}^{T/2} I_0 (x-\Delta x\sin\omega t) {\rm d}t.
\label{eq5}
\end{eqnarray}
Here, $I_0$ ($x$) is the measured electron beam profile without the field according to Fig. \ref{fig4}(c). $\Delta x$ is the deflection amplitude, $\omega$ is the angular frequency of the laser field and $T=2\pi/\omega$ is the laser cycle period (3.4 fs at 1030 nm wavelength). The fit curve (blue, dotted) reproduces the measured deflection profile rather well. The small discrepancy is due to slight remaining inhomogeneity of the optical peak field strength in space and time despite the above mentioned experimental efforts. 

The equations in Section \ref{sec2} predict efficient deflection only for p-polarized light. In Fig. \ref{fig4}(e), we plot the deflection amplitude in $x$- and $y$-directions as a function of the laser's polarization angle, controlled by rotating the half-wave plate; see Fig. \ref{fig4}(a). We see that the deflection along the $x$-axis follows the expected $|\cos\theta_{\rm{pol}}|$ dependency (dotted lines), where $\theta_{\rm{pol}}$ is the polarization angle. On the other hand, we do not observe any significant deflection along the $y$-axis with s-polarized light even at peak laser fields exceeding $10^8$ V/m. This result directly demonstrates the above predicted cancelation of electric and magnetic deflection components after s-polarized light-electron interaction for any electron delay, although the electron momentum is certainly oscillating substantially and in a complex way while propagating through the foil.

Next we considered the achievable peak deflection strengths with different foil materials. Figure \ref{fig4}(f) plots the peak deflection as a function of the laser's peak field strength (in vacuum) for four types of materials (see above). For each foil, there was a direct linear relationship with the applied optical field strength (dotted lines), as expected and observed before\cite{kealhofer2016,Morimoto2017}. This result is a direct evidence that the deflection is indeed induced by the field cycles and not by intensity-based or multi-photon effects such as laser-emitted charge clouds or ponderomotive phenomena. For the Si foil, we observed more than 0.1 mrad deflection, which corresponds to a streaking speed of $>$0.2 mrad/fs at the zero crossing of the deflecting fields. This high streaking speed, which probably exceeds that achievable by metal membranes\cite{kirchner2014,kealhofer2016}, should be beneficial for characterization of ultrashort attosecond pulses; see Section \ref{sec4}. 

In Fig. \ref{fig4}(g) we summarize the slopes of the data in Fig. \ref{fig4}(f) as a function of $d(n-1)$, where $d$ is the thickness and $n$ is the refractive index, a phenomenological parameter responsible for the phase delay between the two half spaces before and behind the foil if disregarding optical interferences (see Appendix \ref{seca2} for details of this approximation). As seen in Sec. \ref{sec2} and Figs. \ref{fig2}(e) and \ref{fig2}(f), the deflection amplitude is almost linear to $d$ and $n$ for small $d$ and $n$. The experimental slopes shown by circles and squares indeed show the linear increase that is expected from the simple theory. Only at large effective foil thicknesses beyond 100 nm, the data shows saturation. This observation agrees with the theoretical results based on Eq. \ref{eq3} denoted with a black curve, which is normalized to the experimental result of 35-nm-thick Si. 

Equations \ref{eq2}-\ref{eq4} also predict a strong influence of the foil angles on the deflection effect. Figures \ref{fig4}(h) and \ref{fig4}(i) show the measured deflection amplitudes as a function of foil angle for 60-nm-thick Si at 0.13 GV/m field strength and for 50-nm-thick $\rm{Si}_3\rm{N}_4$ at 0.17 GV/m, respectively. The laser-electron angle ($\theta_{\rm{laser}}$) was set to $155^\circ$ in these experiments. For both foils, deflection is almost zero at around $\theta_{\rm{foil}} = -10^\circ$, which is close the velocity-matching angle ($-8^\circ$), where zero deflection is expected (see Section \ref{sec2d}). In either direction, the deflection effect increases with angle. The experimental results are reproduced well by the theoretical curves (black) which are normalized to experimental results at $\theta_{\rm{foil}} = 40^\circ$. 

\subsection{Cycle-streaking in Bragg diffraction \label{sec3c}}
\begin{figure}[!tb]
  \includegraphics[width=8.5cm]{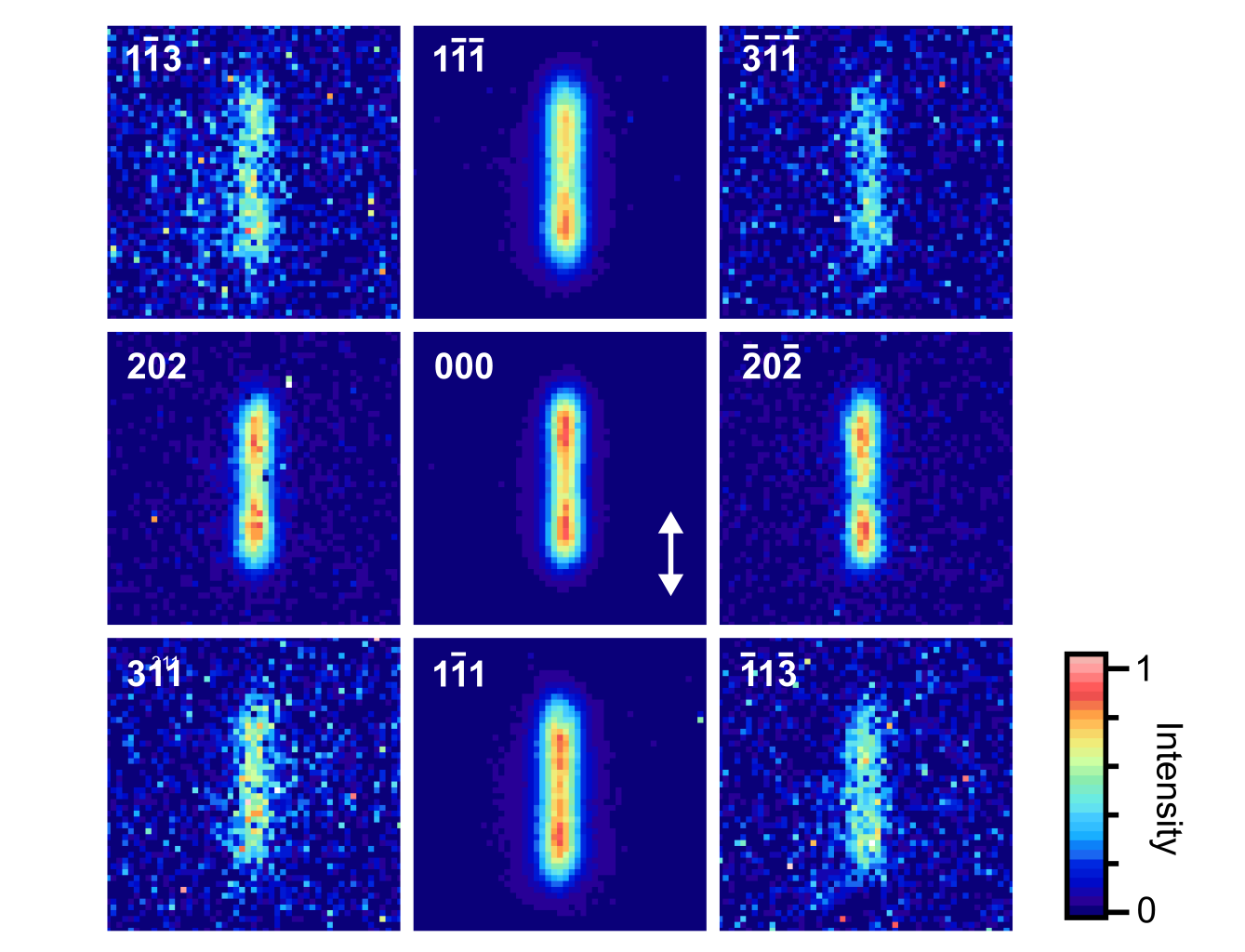}%
  \caption{Attosecond deflection of Bragg spots. The images have a size of $\pm 0.3\times \pm 0.3$ mrad$^2$ and show Bragg diffraction from a single-crystalline Si membrane at [$-1/\sqrt{2}$, $\sqrt{2}$, $1/\sqrt{2}$ ] orientation. The white stripes denote cut-out regions without signal. The Miller indices are shown in each panel. The incident field strength is 0.27 GV/m; the white arrow denotes the polarization direction. \label{fig5}}
\end{figure}

It is also possible to observe the above reported streaking effects in Bragg-diffracted electrons\cite{Morimoto2017}. While in our initial report we had investigated pairs of Bragg spots and the direct beam\cite{Morimoto2017}, we here report on a geometry in which multiple Bragg reflections can be measured at the same time. Figure \ref{fig5} shows several Bragg spots of the single-crystalline 60-nm Si foil if placed at an orientation of [$-1/\sqrt{2}$, $\sqrt{2}$, $1/\sqrt{2}$ ] and fine-tuned to achieve a symmetric pattern with several Bragg spots visible simultaneously. This is possible due to the almost flat Ewald sphere for 70-keV electrons, for which the de Broglie wavelength is 120 times shorter than Si lattice constant, and the finite emittance of the electron beam ($\sim$1 nm)\cite{Morimoto2017}. When the incident laser field is turned on, all the Bragg spots are deflected similarly to the direct beam (000 spot) and produce the same double-lobe pattern at reported for the direct beam in Fig. \ref{fig4}(d). Although the data quality of some Bragg spots is worse than reported before\cite{Morimoto2017}, this result demonstrates the possibility of streaking entire Bragg patterns with attosecond precision.

Bragg-diffracted electrons carry information about lattice constants and the atomic-scale potential within the unit cell\cite{baum2010,shao2010,yakovlev2015}. The sub-cycle streaking of diffracted electrons can therefore potentially be used for imaging light-driven electron dynamics around atoms or nanostructures with sub-optical cycle resolution\cite{baum2010,shao2010,yakovlev2015,shao2017}. At sufficient signal-to-noise, atomic-scale charge displacements will be visible as Bragg-spot intensity changes\cite{morimoto2015,baum2010,shao2010,yakovlev2015,shao2017,stingl2012}. It has also been demonstrated that attosecond-level timing differences of measured Bragg-spot deflections can reveal upper limits on the time it takes to convert a direct-beam electron into a Bragg-diffracted electron by atomic scattering processes in condensed matter\cite{Morimoto2017}. The here demonstrated ability to observe Bragg diffraction and attosecond streaking of multiple Bragg spots in a single experiment should facilitate such endeavours in the future.

\section{EXPERIMENT 2: ATTOSECOND ELECTRON PULSES \label{sec4}}
Electron pulses of shortest possible length in time are a key technology for ultrafast electron diffraction and microscopy for visualizing atomic and electronic motion in space and time\cite{zewail2010,sciaini2011,aseyev2013}. Electron pulses shorter than phonon/vibrational periods\cite{gliserin2015} (tens of femtoseconds) have been achieved by compression concepts with radio-frequency\cite{oudheusden2010,gao2012,chatelain2012,gliserin2012,maxson2017} or THz fields\cite{kealhofer2016}, but seeing light-driven charge carrier dynamics requires attosecond resolution\cite{baum2017ferenc}. Indeed, attosecond electron pulses have recently been achieved in imaging-capable geometries\cite{Morimoto2017,priebe2017} after extensive theoretical\cite{stupakov2001,naumova2004,fill2006,ma2006,sakai2006,baum2007a,dodin2007,kulagin2007,hilbert2009,baum2009,kan2010,liseykina2010,luttikhof2010,andreev2011,hansen2012,li2013,hu2015,lucchio2015,wong2015,greig2016,baumann2017} and experimental studies in relativistic\cite{sears2008,lundh2011,bajlekov2013,guenot2017} and non-relativistic regimes\cite{feist2015,kozak2017,kozaknphys2017}. In our laboratory, we could report some first demonstrations of electron diffraction and microscopy with attosecond resolution\cite{Morimoto2017}. In this section, we describe and discuss a set of theoretical and experimental results on the generation and characterization of attosecond electron pulse trains with different types of dielectric foils. We also report the practical steps that are required for working with attosecond electron pulses in an experiment.

\subsection{Limit of pulse duration \label{sec4a}}
As shown in Section \ref{sec2}, electrons of different arrival time at a foil are accelerated and decelerated periodically with delay time. This modulation at optical frequency reshapes the incoming electron packet into a train of attosecond pulses after subsequent propagation in free space. Each attosecond pulse in the pulse train will be separated by the optical cycle period. 

Before we report the experimental results, let us estimate the shortest possible pulses duration that could be generated with dielectric membranes. Quantum mechanical simulations\cite{baum2017} have revealed a limit of $\Delta t_{\rm{limit}} \approx 2\hbar/(\gamma^3 m_{\rm{e}} v_0 \Delta v_{\rm{max}})$ for the minimum FWHM duration of periodically compressed electron pulses, where $\gamma$ is the relativistic Lorentz factor, $m_{\rm{e}}$ is the electron rest mass and $\Delta v_{\rm{max}}$ is the peak velocity gain or loss from the optical cycles. When $\Delta v_{\rm{max}}$ is negligibly small as compared to $v_0$, this formula can be simplified to $\Delta W_{\rm{max}} \Delta t_{\rm{limit}} \approx 4\hbar$, where $\Delta W_{\rm{max}}$ is the full kinetic energy width associated with $\pm \Delta v_{\rm{max}}$ via $\Delta W_{\rm{max}} = 2 m_{\rm{e}} \gamma^3 v_0  \Delta v_{\rm{max}}$, where $\gamma$ is approximately constant. This energy-time relation agrees with a simple non-quantum-mechanical Fourier analysis that results in  $\Delta W_{\rm{max}} \Delta t_{\rm{limit}} \approx 5\hbar$, (see Appendix \ref{seca4} for details). For 1-ps pulses at 1030 nm wavelength, the laser-damage threshold of a Si foil is $\sim$0.6 GV/m (incident field strength)\cite{morimoto2017-damage}. According to Eq. \ref{eq2} and Fig. \ref{fig2}(c), right panel, we can therefore achieve at suitable angle combinations a maximum an energy broadening of $\Delta W_{\rm{max}} \approx 2 \times 90$ eV = 180 eV. The corresponding quantum limit is $\Delta t_{\rm{limit}} \approx 15$ as, which is considerably shorter than what can be expected with metals or nanostructures due to the limited applicable field strengths there. Electron pulses at tens-of-as duration seem sufficiently short to investigate the entire range of attosecond processes known from optical spectroscopy, but now with the atomic-scale spatial resolution provided by keV-energy free electrons. Even shorter electron pulses could be generated at dielectric membranes by using longer laser wavelengths\cite{baum2017} (see Fig. \ref{fig2}(g)) or shorter laser pulses, for which the damage threshold increases\cite{morimoto2017-damage}. In the compression experiments reported below, the typical amount of energy broadening is $\Delta W_{\rm{max}} \approx$ 5-9 eV and the quantum limit is $\Delta t_{\rm{limit}} \approx$ 290-530 as.

A finite uncorrelated energy spread $\Delta W_{\rm{spread}}$ of the electron beam before the compression can also increase the pulse duration. This energy spread $\Delta W_{\rm{spread}}$ is typically given by the difference between the photon energy of the photoemission laser and cathode material's work function\cite{aidelsburger2010} plus potential technical jitter of the electrostatic acceleration field strength. During the propagation distance $L_{\rm{focus}}$ between the compression membrane and the temporal focus, electron pulses are dispersed. The amount of uncorrelated temporal dispersion $\Delta t_{\rm{spread}}$ in the attosecond pulse train is given by
\begin{eqnarray}
\Delta t_{\rm{spread}} =&& \frac{\Delta v_{\rm{spread}} L_{\rm{focus}}}{v_0^2} \nonumber \\
=&& \frac{2\Delta W_{\rm{spread}}}{\omega \Delta W_{\rm{max}}},
\label{eq6}
\end{eqnarray}
where $\Delta v_{\rm{spread}}$ is the velocity spread of the electron beam caused by the uncorrelated bandwidth $\Delta W_{\rm{spread}}$. $L_{\rm{focus}}$  is a function of the compression strength and the laser frequency as given in Refs.\cite{gliserin2012,baum2017}. 

We find from Eq. \ref{eq6} that higher compression strengths and shorter laser wavelengths reduce the influences of beam imperfections for the achievable pulse duration. The uncorrelated dispersion effect becomes significant if the uncorrelated bandwidth $\Delta W_{\rm{spread}}$ is comparable to or larger than the energy gain $\Delta W_{\rm{max}}$. For our 70 keV electron beams with an estimated $\Delta W_{\rm{spread}} = 0.6$ eV (Ref.\cite{kealhofer2016}) and $\Delta W_{\rm{spread}} \approx$ 5-9 eV, we obtain $\Delta t_{\rm{spread}} \approx$ 70-130 as. This range is lower than the measured pulse duration and also less than the quantum limit (290-530 as), indicating that in our experiment the uncorrelated energy spread is not of importance. We note that typical single-electron pulses are chirped\cite{aidelsburger2010}. Their initial uncorrelated energy spread is therefore redistributed over time and instantaneously lower than directly after photo-emission. The initial energy bandwidth therefore produces in practical cases rather a non-equally-spaced attosecond pulse train instead of lengthening the individual pulses in time. The given  $\Delta t_{\rm{spread}}$ is therefore an upper limit and the actual pulses can be shorter.

\subsection{Experiment for attosecond electron-pulse generation and characterization \label{sec4b}}
\begin{figure*}[!tb]
\includegraphics[width=17cm]{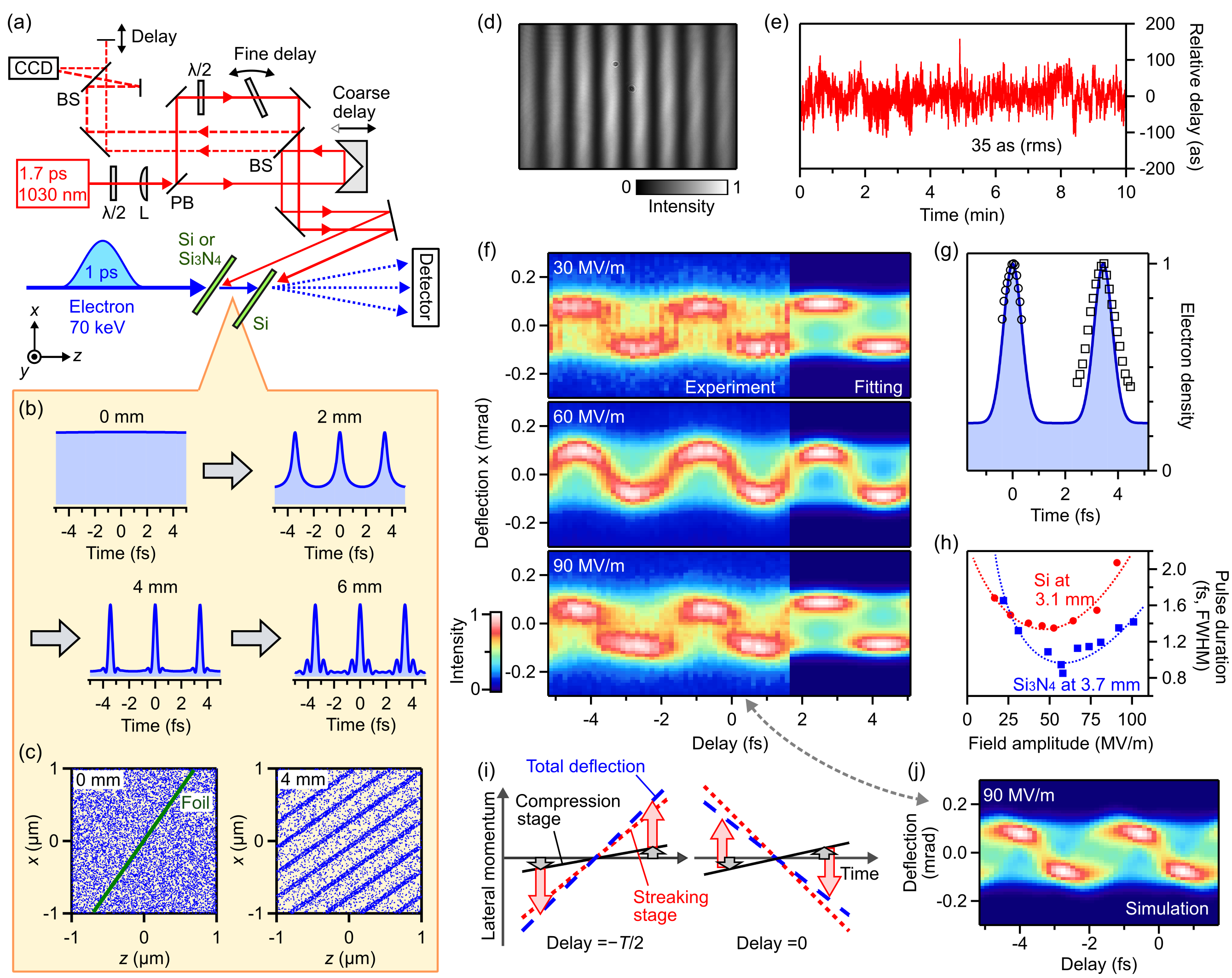}%
\caption{Attosecond electron pulses. (a) Experimental setup. $\lambda/2$, half-wave plate; L, lens; PB, polarizing beam splitter; BS, beam splitter, CCD, auxiliary camera. (b) Time-evolution of the electron density with propagation distance. (c) Snapshots of simulated two-dimensional electron densities. (d) Laser interferogram on the CCD. (e) Interferometric stability of the optical setup. (f) Measured deflectograms at compression strengths of 30 MV/m, 60 MV/m and 90 MV/m. (g) Temporal structure of compressed electron pulses at 60 MV/m. The solid curve is the result of global fitting. Circles and squares represent temporal shapes directly given by the deflectogram (see Section \ref{sec4}). (h) Measured pulse duration as a function of compression field amplitude. Dotted curves are for eye guiding. (i) Space-time couplings and origin of the asymmetry in some data. (j) Deflectogram simulated by including some minor time-dependent deflection at the compression foil. \label{fig6}}
\end{figure*}

Figure \ref{fig6} summarizes the results of attosecond electron pulse metrology with various dielectric foils. The experimental setup is shown in Fig. \ref{fig6}(a). It is based on a sequence of two laser-electron interaction regions with two dielectric membranes (green). The first foil causes compression down to attosecond duration and the second foil is used for pulse metrology by sideways deflection. Like before, picosecond electron pulses were generated by two-photon-photoemission\cite{kasmi2015} by optical pulses form a Yb:YAG laser\cite{schneider2014} and subsequently accelerated to an energy of 70 keV. A magnetic lens (not shown) was applied to produce a weakly converging electron beam (blue) with a divergence of 0.05 mrad and a diameter of $\sim$130 $\upmu$m at the foils (green). For the compression membrane, we applied either a 50-nm thick $\rm{Si}_3\rm{N}_4$ membrane (size, $5\times5$ $\rm{mm}^2$) or a 60-nm-thick Si membrane (size, $3\times3 $ $\rm{mm}^2$), both acquired from Norcada Inc. These foils were oriented at $\theta_{\rm{foil}} = 35^\circ$ and illuminated with a laser field from $\theta_{\rm{laser}} = 155^\circ$ with respect to the electron beam. For the temporal characterization of the compressed attosecond pulses, we used a second foil (green) at some distance, namely a 60-nm-thick Si foil with a size of $0.15\times0.15$ $\rm{mm}^2$ at a distance of 3.7 mm or 3.1 mm from the compressing $\rm{Si}_3\rm{N}_4$ or Si foil, respectively. Both membranes could be aligned and positioned within the vacuum chamber by piezo-driven actuator stages for optimizing the geometry. Beam shapes were detected on the same single-electron detector (phosphor screen plus camera) as before\cite{kealhofer2015a}. 

Figures \ref{fig6}(b) and \ref{fig6}(c) depict theoretical snapshots of the electron pulses during free-space propagation; Fig. \ref{fig6}(b) is obtained from quantum simulations\cite{baum2017} and Fig. \ref{fig6}(c) with point-particles. The electron pulses form a train of attosecond pulses whose duration becomes shortest at 3-4 mm distance. Each attosecond electron pulse is separated by 3.4 fs which corresponds to one optical cycle of 1030-nm light. Due to the non-velocity-matched geometry at the chosen angle combinations, we expect that the electron pulse train is somewhat tilted with respect to the propagation direction (see Appendix \ref{seca5}). Figure \ref{fig6}(c) shows calculated two-dimensional electron densities given by a point-particle simulation with $10^4$ particles. A cut along the $z$-axis gives a classical counterpart to the compressed pulses in Fig. \ref{fig6}(b), but here with additional information on the pulse front tilt. The density at 4 mm distance shows that the compressed electron pulses are tilted by $\sim$$50^\circ$ with respect to the $x$-axis. In the experiment, any such tilt is negligible, because the characterization membrane is parallel to the compression foil; also the two lasers are parallel. This geometry effectively removes any remaining influences of the tilt.

We continue with further explanations of the experiment depicted in Fig. \ref{fig6}(a). The optical beam path (red) consisted of a kind of an interferometer that splits the 1.7-ps laser pulses into two replicas for compression and streaking, respectively. All optics were mounted in rigid mechanical components and placed within a small area of $\sim$15$\times$30 cm$^2$ of footprint. The relative optical phase was monitored by an auxiliary diagnostics part (dashed lines) via an interferometric technique\cite{baum2003} with a beam-profiling camera (CCD) at 30 Hz. An example of the optical interferogram is shown in Fig. \ref{fig6}(d). Figure \ref{fig6}(e) shows the measured fluctuations of the relative phase (or path length) over 10 minutes. We observed a stability of 35 as (rms) without any active stabilization. From the visibility of the fringes we could estimate an upper limit for short-term jitter in the range between 30 Hz (the camera speed) to 25 kHz (Nyquist frequency at 50 kHz laser repetition rate) of $<$300 as (rms). The relative delay between the two pulses was controlled with attosecond precision by rotating a 1-mm-thick fused silica plate inserted into one arm of the interferometer (fine delay in Fig. \ref{fig6}(a)). A single, common lens ($f$ = 600 mm) was used for focusing both pulses to $\sim$300 $\upmu$m diameter ($1/\rm{e}^2$) onto the two membranes with nearly identical wavefronts. Gouy phase differences on the foils were suppressed by placing the two focus locations slightly before the foils. The peak field at the streaking membrane was set to 200 MV/m, while the field strength for compression was set to 10-100 MV/m.

Sometimes we observed in the final attosecond electron streaking data (reported below) a slow interferometric drift that is absent in the auxiliary measurement. The amount was not more than two optical cycles over an hour. This effect can be attributed to drifts of the kinetic energy of the electrons and/or to thermal effects in the laser-excited foils, leading to nanometer-level displacements despite rigid construction of the mounts. In order to account for these effects in the data analysis, we measured each streaking data over a delay of two optical-cycles as a rapid sequence of many fast scans (2-3 minutes), and sorted the data in a post-process by automated Fourier analysis to cancel out the slow drift.

\subsection{Practical requirements for attosecond precision \label{sec4c}}

In principle, deflection of an electron beam by laser cycles like reported in Section \ref{sec3} can serve as an attosecond-resolution metrology of electron pulses\cite{Morimoto2017}, but there are several substantial experimental constraints that need to be considered if aiming for an ultimate time resolution. In Table \ref{table1}, we summarize the most important factors which can affect the resolution. Attosecond streaking works well only when all these conditions are controlled and optimized. 

First, the relative timing between the two laser pulses must be stable at attosecond level. As seen above, our optical setup provides $\sim$35-as (rms) stability without any active stabilization. Second, the drift of the electrons' kinetic energy could, according to Eq. \ref{eq6}, shift the relative timing by $\sim$170 as/eV for membranes separated by 4 mm. Third, displacement jitter of the membranes along the laser beam direction alters the relative timing by $\sim$5 as/nm. Our rigid construction of the two sample holders prevents this effect from becoming significant. Fourth, in case of electron beams with finite diameter, the compression and characterization foils need to be flat, otherwise the relative timing between compression and streaking becomes non-uniform over the beam profile. According to the manufacturer, our dielectric foils have a radius of curvature of $>$10 m, which degrades the timing by only $<$10 as per 100 $\upmu$m of lateral beam size. Fifth, the two foils and the phase fronts of the two laser beams must be parallel to each other, in order to provide the same relative delay along the entire electron beam diameter. This alignment was facilitated by the two independent alignment stages explained above. Sixth, the optical wavefronts (including Gouy phase effects) of the two laser beams should be identical on the two membranes. Thanks to the loose focusing (numerical aperture of $3\times10^{-3}$) through the identical lens and the use of optical focuses slightly before the samples, this effect is negligible. Seventh, the residual divergence of the electron beam causes magnification or demagnification of the tilt of the compressed electron pulses; see Fig. \ref{fig6}(c) and Appendix \ref{seca5} for details. In the experiment, the residual divergence is $\sim$0.05 mrad, causing a tilt mismatch of $\sim0.1^\circ$ and accordingly a blurring of time by $\sim$1 fs for perfectly-parallel membranes. However, tilt magnifications can be compensated for by aligning the foils in a slightly non-parallel way, in order to let the second foil adapt to the given tilt there. Eighth, residual sideways deflections by the compression membrane blur the timing due to different geometrical path lengths between the two foils. This effect is $<$200 as per 0.01 mrad of deflection. Ninth, for meaningful pulse characterization, the streaking speed must exceed one point spread function on the detector per femtosecond. A ten-fold speed can easily be obtained with dielectric membranes (see Section \ref{sec3}). Fluctuations of laser intensity can increase or decrease the streaking speed and therefore blur the pattern, but this effect contributes by less than 30 as per 1$\%$ intensity fluctuation. Fluctuations of the compression strength are almost entirely negligible; see the below discussion of Fig. \ref{fig6}(h). Tenth, elastic and inelastic scattering processes of the electrons inside the foil material might in principle also influence the temporal structure of the streaking data, but such delay effects are below tens of attoseconds\cite{Morimoto2017}.

%
 \begin{table*}[!tb]
 \caption{Practical contributions to temporal mismatch in an attosecond diffraction or microscopy experiment. The values are estimated for 70-keV electrons that are compressed at 4 mm distance by a laser field of 1030-nm wavelength. $z_{\rm{R}}$ is the Rayleigh length of the laser beams. For details about these values see Section \ref{sec4}. \label{table1}}
 \begin{ruledtabular}
 \begin{tabular}{lll}
Factor & Value in the experiment & Contribution to resolution \\
\hline \hline
Stability of interferometer & 35 as (rms, $\leq$30 Hz); & 80 as (FWHM, $\leq$30 Hz); \\
& $<$300 as (rms, $>$30 Hz) & $<$700 as (FWHM, $>$ 30 Hz) \\
E-beam kinetic energy fluctuations & $\sim$1 eV & $\sim$170 as/eV \\
Displacements of foils	& $\sim$nanometers/second & $\sim$5 as/nm \\
Foil flatness	& $\sim$1 nm depth per 100 $\upmu$m lateral size &$<$10 as\\
Parallelism of two foils	 & $<$1 mrad & $<$500 as \\
Laser beams parallelism & $<$1 mrad &$<$200 as \\
Gouy phase differences & $<$0.04 rad & $<$ 20 as \\
Wavefront curvature & $z_{\rm{R}} \sim$ 100 mm &$<$ 100 as \\
Divergence of electron beam	& 0.05 mrad &$<$500 as (after tilt optimization) \\
Streaking speed	 & 0.2 mrad/fs = 17 pixel/fs & 60 as \\
Laser intensity fluctuation & $\sim$2$\%$ (rms) shot-to-shot & $<$60 as \\
Deflection by compression foil & $<$0.02 mrad & $<$400 as \\
Elastic and inelastic scattering & $<$tens of as & negligible
 \end{tabular}
 \end{ruledtabular}
 \end{table*}

\subsection{Generation of attosecond electron pulses \label{sec4d}}
When the compression laser that is incident on the first foil is adjusted to proper field strength, a train of attosecond electron pulses is formed at the location of the second foil\cite{Morimoto2017}. These pulses are in synchrony to the optical cycles there\cite{baum2007a}, and each pulse in the pulse train therefore sees the same deflection dynamics. Measured deflection traces in dependence on phase delay between the two laser beams (so-called deflectograms\cite{kealhofer2016} therefore reveal the electron pulse duration in the pulse train. 

Figure \ref{fig6}(f) shows measured deflectograms of electrons compressed with the $\rm{Si}_3\rm{N}_4$ foil at field amplitudes of 30, 60 and 90 MV/m. In all these deflectograms, localized peaks move up and down with delay time with a period determined by the cycle period of the streaking field (3.4 fs). This oscillation directly demonstrates the presence of the electron pulses shorter than a half optical-cycle. The different shapes of the three deflectograms shown in Fig. \ref{fig6}(f) indicate different pulse durations.
	
We evaluate the duration of the compressed electron pulses by a least-square fitting of each entire deflectogram. When we assume that the electron beam size on the detector is independent on the streaking angles, which is realized by the experimental design, a deflectogram $I(x,\tau)$ is given by
\begin{eqnarray}
&&I(x,\tau) = \int_{-\infty}^{\infty} {\rm d}x' I_0 (x-x') \int_{-T/2}^{T/2 } {\rm d}t \frac{N_{\rm t}}{T} n_{\rm e}(t) \nonumber \\
&& \ \ \ \ \times \delta (x' - \Delta x \sin( \omega t - \omega \tau + \phi_{\rm offset})) \nonumber \\
&& = \frac{N_{\rm t}}{T}  \int_{-T/2}^{T/2 } n_{\rm e}(t) I_0 (x-\Delta x \sin( \omega t - \omega \tau + \phi_{\rm offset}) ) {\rm d}t. \nonumber \\
&&
\label{eq7}
\end{eqnarray}
Here, $\tau$ is the relative delay time between electron and streaking field, $\phi_{\rm{offset}}$ offset is an unknown phase offset accounting for the absolute zero of the delay by the rotating plate, $N_t$ is the number of attosecond pulses in a train, $n_{\rm{e}} (t)$ is the temporal structure of electrons in the attosecond pulses and $I_0 (x)$ is the measured electron beam profile in case of no streaking fields ($\sim$2.3 pixel, rms). We assume for $n_{\rm{e}} (t)$ a Gaussian temporal shape with a rms duration of $\Delta t_{\rm{e}}$ on top of a constant background density $n_{\rm{BG}}$ and a peak density $n_0$, hence $n_{\rm{e}} (t)=n_0  \exp(-t^2/(2\Delta t_{\rm{e}}^2 ))+n_{\rm{BG}}$. The background originates from such electrons that are initially located at non-converging half-cycles of the compression and therefore rather de-compressed than compressed\cite{Morimoto2017,baum2007a,baum2017}. By numerically fitting the four parameters, $\Delta x$, $\phi_{\rm{offset}}$, $\Delta t_{\rm{e}}$ and $n_{\rm{BG}}/n_0$ to the measured two-dimensional deflectograms (2500 data points), we obtain the attosecond electron density $n_{\rm{e}} (t)$ and the pulse duration $\Delta t_{\rm{e}}$. For the deflectogram with $\rm{Si}_3\rm{N}_4$ at 30 MV/m, we obtain a pulse duration of 1320 as (FWHM) or 560 as (rms). At 60 MV/m, we obtain a shorter pulse duration, 820 as (FWHM) or 350 as (rms); see solid curve in Fig. \ref{fig6}(g). At 90 MV/m, we measure 1350 as (FWHM) or 570 as (rms). 

There is another way of analyzing pulse durations in a more direct but less accurate way. We can take only such part of the deflectogram signal that appears at small deflection angles, where the sinusoidal deflection is approximately linear. When we replace $\sin(\omega t-\omega \tau+\phi_{\rm{offset}})$ in Eq. \ref{eq7} by a linear term $\omega t-\omega \tau+\phi_{\rm{offset}}$, the deflectrogram parts around the zero crossings of the deflection directly give the temporal electron pulse profile $n_{\rm{e}} (t)$ convoluted with the electron beam size\cite{kealhofer2016}. The circles in Fig. \ref{fig6}(g) show the result of this analysis around $\tau = 0$ and in the range of $\pm$0.07 mrad for the deflectogram at 60 MV/m. The width of the profile is $\sim$0.9 fs, which is close to the 820 as from the global fit. Collecting vertical profiles around multiple zero crossings gives access to $n_{\rm{e}} (t)$ at wider range of $t$. The squares in Fig. \ref{fig6}(g) show such a profile determined from all of the depicted deflectogram's zero crossings for less than 0.06 mrad deflection angles. This result also agrees with the pulse shape determined by the fitting. Overall, we conclude from this analysis that the global fit with a reasonably complex assumption of pulse shape gives a very reliable and precise value of the pulse duration, thanks to the superior streaking speed that dielectric foils can provide and thanks to over-determination of the data. If a model-free analysis is required, the best approach would probably be a quantum-coherent state reconstruction via energy analysis\cite{priebe2017}, but such an experiment requires a high-resolution energy analyzer, which is non-trivial to construct for laboratory-scale beams at tens-of-keV electron energies or above\cite{gliserin2016}.

Figure \ref{fig6}(h) shows a systematic investigation of the compression dynamics for different foil materials as a function of the compression field strength. The red dots denote results for 50-nm-thick $\rm{Si}_3\rm{N}_4$ and the blue squares show the performance of 60-nm-thick Si. The pulse durations were determined with the fitting procedure. We find that the optimal compression amplitudes are 60 MV/m and 50 MV/m for the $\rm{Si}_3\rm{N}_4$ and Si foil, respectively. At these amplitudes we expect from Eq. \ref{eq2} energy broadenings by $\Delta W_{\rm{max}}=$ 5.0 and 8.7 eV, respectively. With the $\rm{Si}_3\rm{N}_4$  foil, we obtain a minimum pulse duration of 820 as (FWHM), which is close to the quantum limit of 530 as at $\Delta W_{\rm{max}}=$ 5.0 eV. The shortest pulses achieved with the Si foil are slightly longer despite higher energy broadening. This is probably because the experiment was less stable at that time (see Table \ref{table1}).

\subsection{Simultaneous compression and deflection \label{sec4e}}
We turn back to Fig. \ref{fig6}(f) and the three measured deflectograms. The data taken at the highest compression strength of 90 MV/m is not completely symmetric around the half cycles at $\tau= 0$ and $-1.7$ fs, although in principle the streaking physics should be the same at these two points, just with opposite direction. In energy-domain attosecond streaking of photoelectrons, such asymmetry is a sign of chirp (time-energy correlation) of the attosecond pulses\cite{hofstetter2011}. In contrast, in our sideways deflection experiment, the origin of the asymmetry is a time-dependent angular dispersion that occurs simultaneously with the acceleration/deceleration at the compression foil. From Eq. \ref{eq3} and the absence of spot width increases of the direct beam after compression\cite{Morimoto2017}, we conclude that this effect is rather small ($\leq$0.02 mrad), but nevertheless measurable in the streaking for cases of over-compression. 

Figure \ref{fig6}(i) depicts the mechanism of the asymmetry. As discussed in Section \ref{sec2}, electrons can obtain both a time-dependent forward/backwards and sideways momentum simultaneously at the compression membrane, depending on the angle combination. If the compression strength is below or above the optimum value, the electron pulses are under-compressed or over-compressed at the target and therefore chirped (time-energy correlation). At the given angle combination, the electrons have also a slight sideways momentum that changes over the pulse duration (black lines). This sideways momentum can be assumed to vary linearly with time. The characterization membrane adds a second, much stronger time-dependent lateral momentum (red dotted lines) on purpose ($>$0.1 mrad). The sideways momentum enhances or reduces the overall deflection (blue dashed lines) depending on the delay $\tau$ and specifically in a different ways for a rising or a falling zero-crossing. This effect causes asymmetric deflectogram shapes, as seen, for example, in the experiment at 90 MV/m.

Figure \ref{fig6}(j) shows a deflectogram that was simulated by assuming a sideways momentum change by 0.03 mrad/fs in the time range where the attosecond electron pulses have finite intensity. The simulated deflectogram well reproduces the measured asymmetry for the case of 90 MV/m; compare Fig. \ref{fig6}(f), third panel. In principle, the deflectogram at 30 MV/m compression strength is also asymmetric, but the effect is too weak to be observed. In practice, the time-dependent deflections of the compression stage can either be made negligible by choosing a far enough locations of the temporal focus, or alternatively by using a better angle combination than in the reported experiment, namely a point on Figs. \ref{fig2}(c) and \ref{fig2}(d) where there is no deflection and only compression. Interestingly and favourably, such condition is compatible with velocity matching, as discussed above.

\section{THEORETICAL RESULTS ON ABSORBING MATERIALS \label{sec5}}
\begin{figure*}[!tb]
\includegraphics[width=17cm]{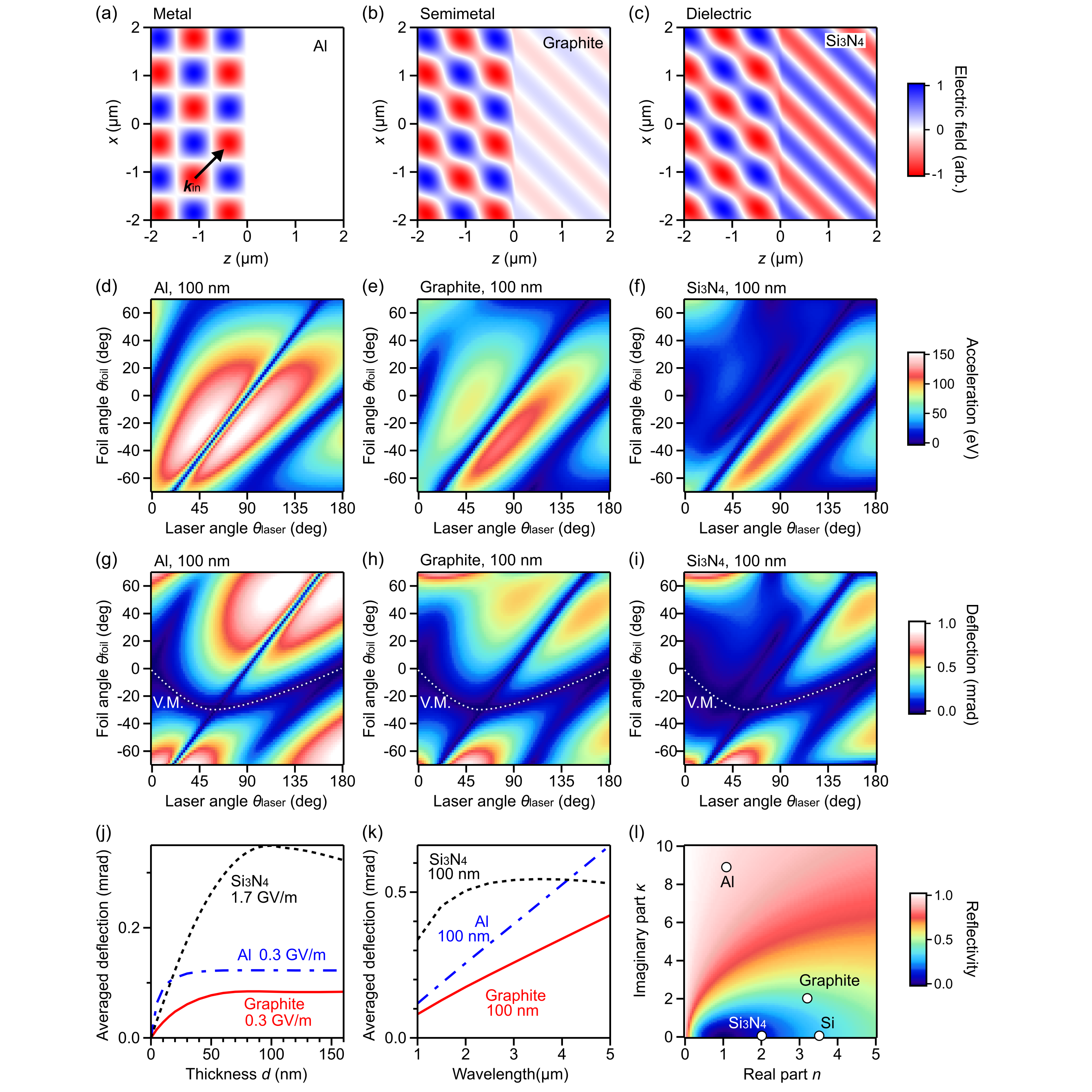}%
\caption{Comparison of metallic, absorbing, and dielectric foils. (a)-(c) Time-frozen snapshots of p-polarized electric fields around 100-nm-thick aluminum, graphite and Si$_3$N$_4$ membranes located at $z=0$. (d)-(f) Peak acceleration for 70-keV electrons. (g)-(i) Peak deflection. The white dotted curves represent configurations with velocity-matching. (j) Mean deflection as a function of thickness for measured damage thresholds. (k) Mean deflection as a function of laser wavelength, also for measured damage thresholds. (l) Reflectivity of materials as a function of the real and imaginary part of the refractive index $n+{\rm{i}}\kappa$. The reflectivity is given for 1030-nm light and normal incidence. \label{fig7}}
\end{figure*}

Not only metal mirrors\cite{kirchner2014,kealhofer2016,vanacore2017pre,plettner2005} or dielectrics\cite{Morimoto2017} can be applied for all-optical electron control, but there is also the possibility to use partially absorbing materials\cite{priebe2017}, in which part of the optical energy is neither reflected nor transmitted. In this section, we report theoretical results for absorbers and compare them with those of metals and dielectrics. Our theory of Section \ref{sec2} is indeed also applicable to materials with complex refractive indices $n + \mathrm{i}\kappa$. Examples of such materials are narrow-gap semiconductors, semi-metals or metals. We choose the example materials, graphite ($n= 3.2$, $\kappa= 2.0$) and aluminum ($n=1.1$, $\kappa=8.9$), because such foils have been investigated before\cite{priebe2017,kirchner2014,kealhofer2016}. Figure \ref{fig7} summarizes the results; we assume 1030-nm laser excitation. Figures \ref{fig7}(a) and \ref{fig7}(b) depict time-frozen snapshots of the electric fields around the Al and graphite membranes, respectively. For comparison, Fig. \ref{fig7}(c) shows the fields around a $\rm{Si}_3\rm{N}_4$ foil. The thickness of all foils is 100 nm. We see that the transmitted fields are very weak for graphite and almost zero for Al, while $\rm{Si}_3\rm{N}_4$ transmits almost everything. For materials with finite $\kappa$, the surface reflectivity is enhanced and the electromagnetic fields inside the membranes decay exponentially with the distance from the surface. The field penetration depths, characterized by $c/\omega \kappa$, are 18 nm for aluminum and 82 nm for graphite. The plotted field distributions clearly illuminate the differences of the three regimes and main effects: complete reflection for metal mirrors, optical phase shift for dielectrics and both for absorbers.

Figures \ref{fig7}(d)-(i) show the peak acceleration and deflection for the three types of membranes for all angle combinations. As above, the kinetic energy of electron is 70 keV and we assumed a p-polarized field of 1030 nm wavelength and 1 GV/m peak field strength. We could not find any deflections with s-polarized light for the all three materials. The details of our analytical approach are given in Appendix \ref{seca3}. The results reveal a general trend. From metal over graphite to dielectric there is more structure and less symmetry around the diagonal line $\theta_{\rm{foil}} \approx \theta_{\rm{laser}}-90^\circ$ (laser grazing incidence). In metals there are negligible thin-film interferences and optical phase shifts that could critically depend on angles, but such interferences increasingly arise for in graphite and $\rm{Si}_3\rm{N}_4$, despite the large thickness of 100 nm. The overall magnitudes of acceleration and deflection are almost the same; remember the assumption of 1 GV/m, although the actual damage threshold is different (see below). Figures \ref{fig7}(g)-(i) reveal that the deflection by any material is zero at the velocity matching condition (white dotted curves). This zero deflection is due to the contribution of magnetic fields and has important consequences for pulse compression and metrology; see below in Section \ref{sec6}.

In order to further elucidate the physics in the three regimes, we report dependencies on thickness and laser wavelength. Figure \ref{fig7}(j) shows the predicted peak deflection (averaged over all angles like in Section \ref{sec2}) as a function of foil thickness for the three cases. For practical purposes, we scaled the plotted peak deflections to the laser damage threshold of the three materials. According to experiments\cite{morimoto2017-damage}, the damage threshold of free-standing $\rm{Si}_3\rm{N}_4$ membranes is 1.7 GV/m. Damage thresholds of Al and graphite membranes are approximated by the measured values for Cu and graphene (0.3 GV/m for both). In Fig. \ref{fig7}(j), Al (blue curve) and graphite (red curve) give almost constant deflection strength at large thickness, because the absorption depth (tens of nm) dominates the interaction for thicker foils. On the other hand, the deflection by $\rm{Si}_3\rm{N}_4$ (black curve) continuously changes with thickness because of continuing thin-film interferences (see Section \ref{sec2}). 

Figure \ref{fig7}(k) elucidates the dependency on laser wavelength. We assumed constant damage thresholds between 1-5 $\upmu$m like indicated by experiments on dielectrics\cite{simanovskii2003,grojo2013} and metals\cite{porteus1981}. Graphite has nearly constant absorbance\cite{nemanich1977,bandara2017} and therefore constant damage thresholds can be expected. Figure \ref{fig7}(k) shows that the Al foil (blue line) provides an effect that is linear to wavelength\cite{kirchner2014,andrey_thesis2017}. This linearity is also evident from Eqs. \ref{eq2}-\ref{eq4}. In the dielectric (black dotted line), the effect at increasing wavelength is in part cancelled by the simultaneous reduction of the optical phase shift (see Section \ref{sec2}). Graphite (red line) has almost no phase effects because of its strong absorption. Figure \ref{fig7}(l) summarizes the considered materials in terms of $n$, $\kappa$ and reflectivity. Overall, metallic membranes are ideal at long wavelengths like in the THz regime\cite{kealhofer2016}, but dielectric foils are ideal for visible/near-infrared wavelengths and therefore for the attosecond regime\cite{Morimoto2017,priebe2017}. 

\section{CONTRIBUTION OF MAGNETIC FIELDS \label{sec6}}
Before concluding this work, we discuss the role of magnetic fields, which are typically neglected in earlier studies with nanostructures\cite{barwick2009,yurtsever2012}. Generally, at electron velocities approaching a substantial fraction of the speed of light, for example 0.5 at 70 keV, magnetic and electric components of the laser fields play a comparable role. 

\begin{figure*}[!tb]
\includegraphics[width=17cm]{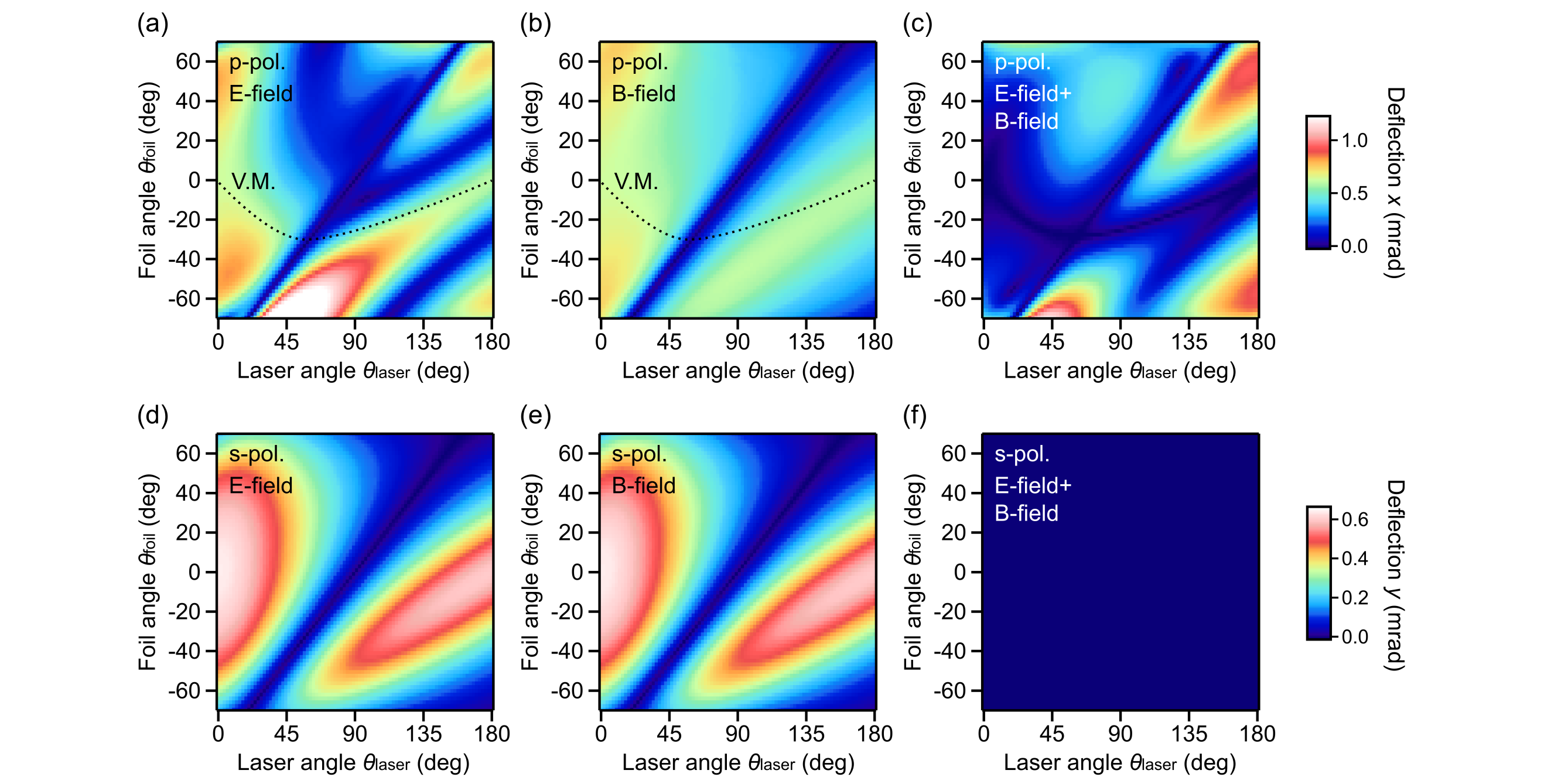}%
\caption{Effect of magnetic fields. (a)-(c) Deflection $\Delta p_x$ with p-polarized light and a 60-nm-thick Si foil. (a) Electric part only. (b) Magnetic part only. (c) Complete effect. Dotted curves V.M. represent the electron-laser velocity-matching condition. (d)-(f) Deflection $\Delta p_y$ with s-polarized light. (d) Electric part only. (e) Magnetic part only. (f) Complete effect. All simulations are for 60-nm-thick Si and 1 GV/m laser fields at 1030 nm wavelength. \label{fig8}}
\end{figure*}

Figure \ref{fig8} shows the magnetic and electric contributions individually, using the case of a 60-nm-thin Si foil. First, we discuss the deflection $\Delta p_x$ in p-polarized fields (compare the experiment in Section \ref{sec3}). Figure \ref{fig8}(a) shows the electric part of the effect, Fig. \ref{fig8}(b) shows the magnetic part, and Fig. \ref{fig8}(c) shows the overall effect. The electric and magnetic contributions have comparable magnitude but very different angle dependencies; none of them can approximate the overall result. Dotted curves marked as V.M. represent the velocity-matching condition; electric and magnetic deflections are substantial but cancel out. Second, we consider the deflection $\Delta p_y$ by s-polarized light. Figures \ref{fig8}(d)-(f) show again the electric and magnetic contributions and the overall result. We find that the electric and the magnetic contributions are both substantial, but the combination becomes zero at all laser/foil angles, as shown in Fig. \ref{fig8}(f). The two cases, zero-deflection at velocity-matching and zero-deflection for s-polarization, are also found for metallic and absorbing membranes. The mechanism seems universal for all types of membranes. Importantly, although the net effect might be zero, electron momentum does oscillate considerably while being inside the laser fields and inside the membrane. These results show clearly the importance of magnetic field components, and suggest that dynamical magnetic effects can indeed be studied by experiment, namely by using waveform electron microscopy at different electron velocities\cite{Ryabov2016}. 

\section{GENERAL FINDINGS AND OUTLOOK \label{sec7}}
Dielectric and absorbing foils are a valuable alternative to nanostructures or metal mirrors for controlling electron beams and wave packets with sub-optical-cycle precision and therefore on attosecond time scales. Various conditions can be found to shape the interplay of deflection and acceleration/compression and their relative strengths or signs for the desired application. Dielectrics generally have the highest possible laser damage thresholds and are therefore an ideal third body for laser-electron control at visible/near-infrared wavelengths. They offer acceleration/deceleration by hundreds of eV per laser cycle, which allows the all-optical formation of electrons pulses as short as 10 as or below.

The above set of experimental results in comparison to calculations demonstrates that a simple, classical theory based on electromagnetic waves can fully and entirely reproduce all kinds of deflection and compression experiments with any kind of membranes for almost all conceivable geometries. No photons are needed for understanding the physics or explaining the results. In many cases, the magnetic and electric components of the laser wave play a comparable role and cause two non-trivial cancelations that are useful for experiments. Attosecond electron diffraction and microscopy\cite{Morimoto2017} have potential to investigate electrically induced dynamics, as predicted\cite{baum2010,shao2010,yakovlev2015} and indicated by experiment\cite{morimoto2015,stingl2012}, but magnetic effects are also substantial can be revealed in materials and nanostructures. The reported experimental details and procedures for all-optical attosecond control of electrons will help proliferating free-electron-based attosecond science as a novel tool for investigating light-matter interaction on its natural space-time dimensions.

\begin{acknowledgments}
This work was supported by the European Research Council and the Munich-Centre for Advanced Photonics. YM acknowledges support from a JSPS Postdoctoral Fellowship for Research Abroad. We thank Simon Stork for sample handling, Maxim Tasrev for discussions, Ferenc Krausz for general support and Bo-Han Chen, Andrey Ryabov and Dominik Ehberger for help with the laser.
\end{acknowledgments}

\appendix
\section{ANALYTICAL FORM OF ELECTROMAGNETIC FIELDS \label{seca1}}
Equations \ref{eq2}-\ref{eq4} of the main text allow to predict the momentum gain of an electron from a laser field if interacting with a non-absorbing dielectric membrane of arbitrary thickness and material. Evaluation requires knowledge of the electric and magnetic fields outside and inside a foil. While the transmitted intensity or reflectivity of thin layers has been studied a lot, we find it useful to report here analytically the electric and magnetic fields including the space-time propagation of the optical cycles, because those are decisive for the attosecond physics of compression and deflection. For the cases of absorbing materials and metals, see also Appendix \ref{seca3}. 

\begin{figure}[!tb]
\includegraphics[width=8.5cm]{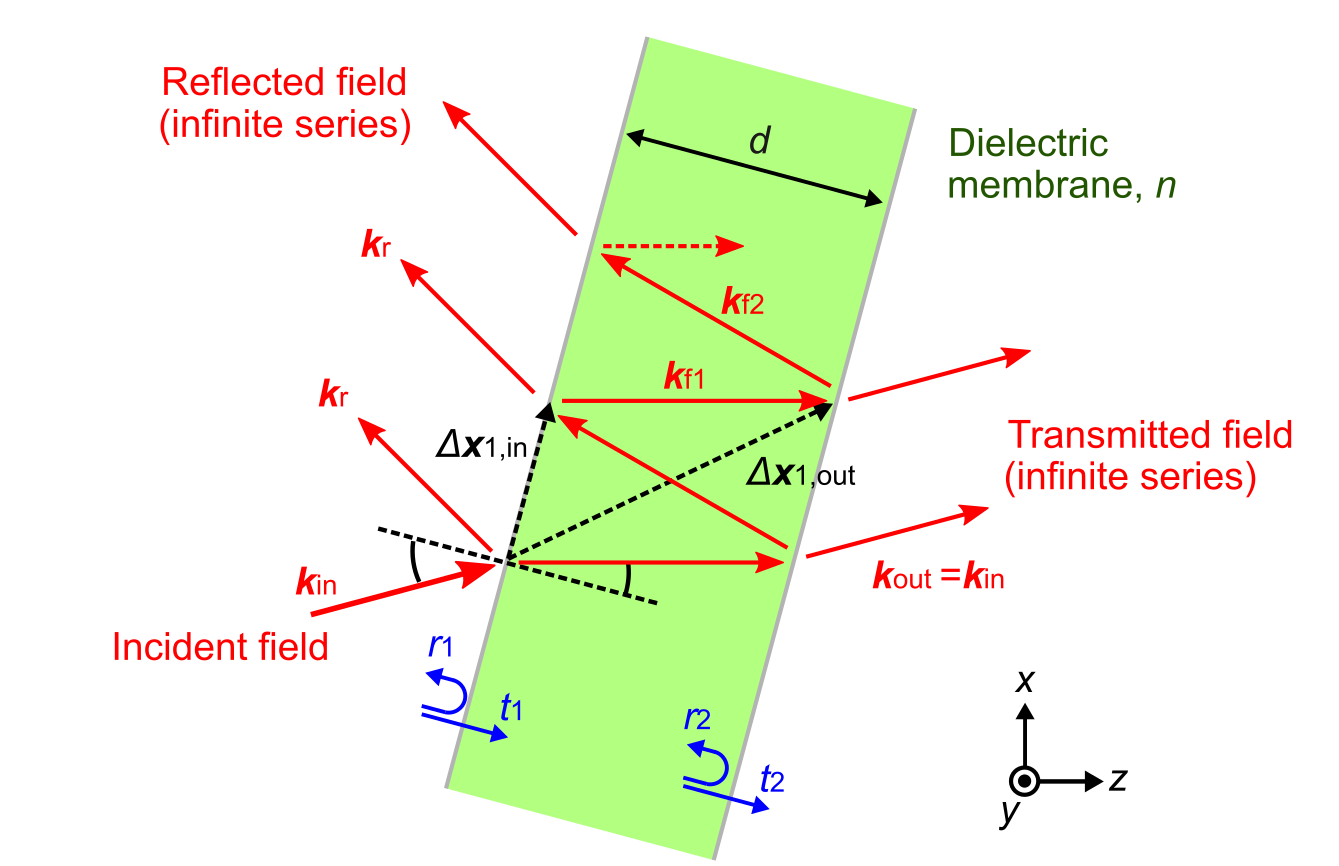}%
\caption{Calculation scheme for obtaining electromagnetic field cycles inside and outside of a dielectric membrane (green). See Appendix \ref{seca1} for definitions of the position and wave vectors $\Delta \boldsymbol{x}$, $\boldsymbol{k}$ and the relative field amplitudes $r, t$. \label{fig9}}
\end{figure}

We consider an incident plane wave of wave vector ${\boldsymbol{k}}_{\rm{in}}$ and monochromatic frequency $\omega$ that converts in the foil to an infinite series of monochromatic plane waves by multiple reflections from the foil surfaces; see Fig. \ref{fig9}. The electric field in vacuum on the laser-incident side (left in Fig. \ref{fig9}) is given by
\begin{eqnarray}
&&E(\boldsymbol{x},t) = \boldsymbol{\epsilon}_{\rm in} E_0 \cos(\boldsymbol{k}_{\rm in} \boldsymbol{x} - \omega t) + \boldsymbol{\epsilon}_{\rm r} r_1 E_0 \cos(\boldsymbol{k}_{\rm r} \boldsymbol{x} - \omega t) \nonumber \\
&& \ \ \ + \boldsymbol{\epsilon}_{\rm r} t_1 t_2 E_0 \sum_{m=1}^{\infty} r_2^{2m} \cos(\boldsymbol{k}_{\rm r} (\boldsymbol{x} - \boldsymbol{x}_{m, {\rm in}}) - \omega t + 2m\phi_{\rm f}), \nonumber \\
&&
\label{eqa1}
\end{eqnarray}
where $\boldsymbol{\epsilon}_{\rm{in}}$ and  $\boldsymbol{\epsilon}_{\rm{r}}$ are polarization vectors for incident and reflected lights, $E_0$ is electric field amplitude of the incident light, ${\boldsymbol{k}}_{\rm{in}}$ and ${\boldsymbol{k}}_{\rm{r}}$ are wave vectors of incident/reflected lights and $2\phi_{\rm{f}} = 2\omega nd/c \cos \theta_{\rm t}$  is the phase delay due to one round trip at the refraction angle of $\theta_{\rm t}$. The coefficients $r_1$ and $t_1$ are relative amplitudes for reflected and transmitted waves for a light travelling from vacuum into the foil; $r_2$ and $t_2$ are similar for the transition from the foil to vacuum. $\Delta \boldsymbol{x}_{m,\rm{in}}$ is the shift in position due to a round trip in the foil, and is given by $\Delta \boldsymbol{x}_{m,\rm{in}}=(2md \tan\theta_{\rm t},0,0)$ if the foil surface is parallel to $x$-axis. The first term on the right-hand side of Eq. \ref{eqa1} represents the incident wave while the second and the third terms denote the reflected waves. In a similar way, the electric field in vacuum on the laser-outgoing side (right in Fig. \ref{fig9}) is given by
\begin{eqnarray}
&& E(\boldsymbol{x},t) = \boldsymbol{\epsilon}_{\rm in} t_1 t_2 E_0 \sum_{m=0}^{\infty} r_2^{2m} \nonumber \\
&& \times \cos(\boldsymbol{k}_{\rm in} (\boldsymbol{x} - \Delta \boldsymbol{x}_{m, {\rm out}}) - \omega t + (2m+1)\phi_{\rm f}),
\label{eqa2}
\end{eqnarray}
where $\Delta \boldsymbol{x}_{m,\rm{out}}$ is a shift in wave origins and is given by $\Delta \boldsymbol{x}_{m,\rm{out}} =((2m+1)d\tan\theta_{\rm t},0,d)$ when foil surface is parallel to $x$-axis. Finally, the electric field inside the dielectric foil is given by
\begin{eqnarray}
E(\boldsymbol{x},t) &=& \boldsymbol{\epsilon}_{\rm f1} t_1 E_0 \sum_{m=0}^{\infty} r_2^{2m} \nonumber \\
&& \ \times \cos(\boldsymbol{k}_{\rm f1}( \boldsymbol{x} - \boldsymbol{x}_{m, {\rm in}}) - \omega t +2m\phi_{\rm f})  \nonumber \\
&&+\boldsymbol{\epsilon}_{\rm f2} t_1 E_0 \sum_{m=0}^{\infty} r_2^{2m+1} \nonumber \\
&& \ \times \cos(\boldsymbol{k}_{\rm f2}( \boldsymbol{x} - \boldsymbol{x}_{m, {\rm out}}) - \omega t + (2m+1)\phi_{\rm f}), \nonumber \\
&& 
\label{eqa3}
\end{eqnarray}
where ${\boldsymbol{\epsilon}}_{\rm{f1}}$ and ${\boldsymbol{\epsilon}}_{\rm{f2}}$ are polarization vectors for forwards and backwards waves, respectively; ${\boldsymbol{k}}_{\rm{f1}}$ and ${\boldsymbol{k}}_{\rm{f2}}$ are their wave vectors. Magnetic fields are obtained in a similar way.

Time integrals of these electromagnetic fields, as required in Eqs. \ref{eq2}-\ref{eq4} of the main text, can easily be analytically evaluated thanks to this plane-wave expansion. An integral of an electric or a magnetic field in the general form of ${\boldsymbol{\epsilon}}F \cos(\boldsymbol{k}\boldsymbol{x}-\omega t+\phi)$ with $\boldsymbol{x}=(0,0,v_0 (t-t_0))$ is given by
\begin{eqnarray}
&&\int_{t_a + t_0}^{t_b + t_0}  {\boldsymbol{\epsilon}}F \cos(\boldsymbol{k}\boldsymbol{x}-\omega t+\phi) {\rm d}t \nonumber \\
&& = \frac{ {\boldsymbol{\epsilon}}F}{\omega - k_z v_0} \Bigr [ \sin(\omega t_0 + (\omega - k_z v_0)t - \phi) \Bigl ]_{t_a}^{t_b},
\label{eqa4}
\end{eqnarray}
where $k_z$ is $z$-component of the wave vector $\boldsymbol{k}$. This general form can be applied for the integrals of all the electromagnetic fields before, inside and after a dielectric membrane. 

We quickly discuss the dependence of the overall momentum change with $t_0$, the time of arrival at the foil. Equation \ref{eqa4} is a sinusoidal function of $\omega t_0$. The overall momentum change (Eqs. \ref{eq2}-\ref{eq4}) is a sum of many such contributions. Because any sum of sinusoidally oscillating functions at the same frequency remains sinusoidally oscillating, the overall momentum gain is therefore a sinusoidal function of $\omega t_0$. The complexity of the optical interferences only changes the magnitude and phase of the effect, but not its sinusoidal dependence in time. 

\section{BREWSTER'S ANGLE \label{seca2}}
\renewcommand{\theequation}{A\arabic{equation}}
\setcounter{equation}{4}

We found in Fig. \ref{fig3} that the optical phase shift between the light waves on both sides of a dielectric membrane play the major role for acceleration and deflection, while the fields inside the membrane are less important, see Section \ref{sec2d}. Here, we discuss the special case of Brewster's angle, in which this picture can be put on solid ground. Although we consider acceleration here, essentially the same conclusions can also be drawn for deflection. 

We can express $\Delta p_z$ at the Brewster's angles in an explicit form. By substituting Eqs. \ref{eqa1}-\ref{eqa3} with $t_1=1/n$, $t_2=n$ and $r_1=r_2=0$ into Eq. \ref{eq2}, we obtain 
\begin{eqnarray}
&&\Delta p_z (t_0) =  \nonumber \\
&& \ \ \frac{2eE_0 \epsilon_{{\rm in},z}}{\omega - k_{{\rm in},z}v_0} \sin(\pm \frac{\phi_{\rm f}}{2} - \frac{\omega - k_{{\rm in},z}v_0}{2}T_{\rm f}) \cos(\omega t_0 - \frac{\phi_{\rm f}}{2})  \nonumber \\
&& \ \  -  \frac{2eE_0 \epsilon_{{\rm in},z}}{n(\omega - k_{{\rm f1},z}v_0)} \sin(- \frac{\phi_{\rm f}}{2} - \frac{\omega - k_{{\rm f},z}v_0}{2}T_{\rm f}) \cos(\omega t_0), \nonumber \\
&&
\label{eqa5}
\end{eqnarray}
where $\epsilon_{\rm{in},z}$, $\epsilon_{\rm{f1},z}$, $k_{\rm{in},z}$ and $k_{\rm{f1},z}$ are the $z$-component of $\boldsymbol{\epsilon}_{\rm{in}}$, $\boldsymbol{\epsilon}_{\rm{f1}}$, $\boldsymbol{k}_{\rm{in}}$ and $\boldsymbol{k}_{\rm{f1}}$, respectively. $\phi_{\rm{f}} =\omega nd/c \cos\theta_{\rm{t}}$ gives the optical phase shift in the absence of thin-film interference and $T_{\rm{f}}=d/v_0 |\cos\theta_{\rm{foil}} |$ is the duration (typically tens of attoseconds) that the electron spends inside of the foil. The $\pm$ appearing in the first term on the right-hand side is positive when electrons and laser hit the membrane from the same side, and negative for different incidence sides. The first term on the right-hand side gives the contribution from the fields outside the membrane while the second term gives the contribution of the fields inside. Compared to the first term, the second term is reduced by the refractive index $n$ and is therefore substantially smaller in most cases, especially for highly refractive materials (compare Section \ref{sec2}).

The optical phase shift $\phi_{\rm{f}}$ depends on the thickness $d$, laser frequency $\omega$ and refractive index $n$. Using Eq. \ref{eqa5}, we discuss the dependence of peak acceleration on these parameters. First, if we set $n=1$ (no foil), we get $\Delta p_z (t_0 )=0$ because the first and the second terms are cancelled out. With increasing $n$, the difference between the first and the second term increases. Accordingly, the peak acceleration also increases, approximately proportionally to $(n-1)$. Second, at $d=0$ (no foil), we also obtain $\Delta p_z (t_0 )=0$ because $T_{\rm{f}}$ and $\phi_{\rm{f}}$ become zero. For simplicity of the following discussions, we neglect the second term on the right hand side of Eq. \ref{eqa5}; this neglects the fields inside the foil. Under this approximation, $\Delta p_z$ is directly proportional to $\sin(\Delta \phi/2)$, where the phase $\Delta \phi =\pm \phi_{\rm{f}} - (\omega- k_{\rm{in},z} v_0 ) T_f$ is the phase jump that electrons feel when passing though the membrane (see below). If $|\Delta \phi|$ is small enough to give $\sin(\Delta \phi/2)\approx \Delta \phi/2$, the amplitude of $\Delta p_z$ becomes $eE_0 \epsilon_{\rm{in},z} \Delta \phi/(\omega-k_{\rm{in},z} v_0)$. Because $\Delta \phi$ is linear with $d$, the peak acceleration is directly proportional to $d$. Third, recalling that $\phi_{\rm{f}}$, $\Delta \phi$ and $k_{\rm{in},z}$ are proportional to $\omega$, we find that the acceleration amplitude $eE_0 \epsilon_{\rm{in},z} \Delta \phi/(\omega-k_{\rm{in},z} v_0)$ becomes independent of $\omega$. In other words, the $1/\omega$-dependence coming from the integrals of the fields is cancelled by $\omega$-dependence of the phase $\Delta \phi$. Accordingly, the peak acceleration in dielectrics at Brewster's angle is independent of the laser wavelength. 

We quickly explain why $\Delta \phi$ as defined above is indeed the phase shift that the electrons see between entering and leaving the foil. If electron exits the first field $E_1 (\boldsymbol{r},t)=E_0 \cos(\boldsymbol{k} \boldsymbol{r}-\omega t)$ at $t=-T_{\rm{f}}/2$ from $\boldsymbol{r}=(0,0,-v_0 T_{\rm{f}}/2)$ and enters the second field $E_2 (\boldsymbol{r},t)=E_0 \cos(\boldsymbol{k} \boldsymbol{r}-\omega t \pm \phi_{\rm{f}})$ at $t=T_{\rm{f}}/2$ into $\boldsymbol{r}=(0,0,v_0 T_{\rm{f}}/2)$, the phase difference the electron feels is given by $k_z v_0 T_{\rm{f}}-\omega T_{\rm{f}} \pm \phi_{\rm{f}}$, which is identical to $\Delta \phi$.

We return to the expression of $\Delta \phi =k_z v_0 T_{\rm{f}}-\omega T_{\rm{f}} \pm \phi_{\rm{f}}$. Using the relationship, $|k_z v_0 | \leq \omega v_0/c \leq \omega$, valid for fields in vacuum, the sum of the two terms $k_z v_0 T_{\rm{f}} -\omega T_{\rm{f}}$, which corresponds to the phase jump due to the time delay of the electron passing through the foil, is always negative. This indicates that when the sign in front of $\phi_{\rm{f}}$ is also negative, the two effects in $\Delta \phi$ are added up, and the peak acceleration is therefore enhanced. This is the case when the electron and laser hit the membrane from different sides (counter-propagation). Indeed, Figs. \ref{fig2}(c) and \ref{fig2}(d) show that the peak accelerations are generally stronger at the lower-right region ($ \theta_{\rm{foil}} < \theta_{\rm{laser}}-90^\circ$), where electrons and laser come from different sides of the membrane, as compared to the upper-left part ($ \theta_{\rm{foil}} > \theta_{\rm{laser}}-90^\circ$), where electrons and laser hit the same side of the membrane.

\section{ELECTROMAGNETIC FIELDS AT ABSORBING MATERIALS \label{seca3}}
\renewcommand{\theequation}{A\arabic{equation}}
\setcounter{equation}{5}

Here we quickly outline a procedure for generalizing the above Eqs. \ref{eqa1}-\ref{eqa3} to materials with a complex refractive index $n+\rm{i}\kappa$. This generalization allows us to predict the acceleration and deflection at almost any kind of membrane including metals, semi-metals and narrow-bandgap materials. First, we replace the electric fields in the form of $\boldsymbol{\epsilon} F \cos(\boldsymbol{k}\boldsymbol{x}-\omega t+\phi)$ by ${\rm{Re}} \bigl \{ \boldsymbol{\epsilon} F {\rm{e}}^{\mathrm{i} \boldsymbol{k}\boldsymbol{x}-\mathrm{i}\omega t+\mathrm{i}\phi} \bigr \} $. The wave vector $\boldsymbol{k}$ is complex $\boldsymbol{k}=\boldsymbol{k}^{\rm{R}}+{\rm{i}} \boldsymbol{k}^{\rm{I}}$ inside memrbanes, and the imaginary part $\boldsymbol{k}^{\rm{I}}$ contributes to the decay of field amplitudes. The amplitude $F$, which is given by the coefficients $r_1$,$r_2$,$t_1$ and $t_2$, is also a complex value. Below, we give these coefficients and wave vectors by referring to the more elaborate, comprehensive discussions in Ref.\cite{fowlesbook}. The relative amplitudes of the reflected and transmitted waves at the vacuum-foil boundary are given by
\begin{eqnarray}
r_1 = \frac{ -(n+{\rm i}\kappa)\cos\theta_{\rm in} + \cos\Theta_{\rm t}}{(n+{\rm i}\kappa)\cos\theta_{\rm in} + \cos\Theta_{\rm t}},
\label{eqa6}
\end{eqnarray}
\begin{eqnarray}
t_1 = \frac{ 2\cos\theta_{\rm in} }{(n+{\rm i}\kappa)\cos\theta_{\rm in} + \cos\Theta_{\rm t}},
\label{eqa7}
\end{eqnarray}
for p-polarized light, and 
\begin{eqnarray}
r_1 = \frac{ \cos\theta_{\rm in} - (n+{\rm i}\kappa) \cos\Theta_{\rm t}}{\cos\theta_{\rm in} + (n+{\rm i}\kappa) \cos\Theta_{\rm t}},
\label{eqa8}
\end{eqnarray}
\begin{eqnarray}
t_1 = \frac{ 2\cos\theta_{\rm in} }{\cos\theta_{\rm in} + (n+{\rm i}\kappa) \cos\Theta_{\rm t}},
\label{eqa9}
\end{eqnarray}
for s-polarized light. Here $\theta_{\rm{in}}$ is the laser incident angle with respect to foil normal, see Fig. \ref{fig9}. The complex refraction angle $\Theta_{\rm{t}}$ is given by
\begin{eqnarray}
\cos\Theta_{\rm t}= \sqrt{ 1- \frac{\sin^2 \theta_{\rm in}}{ (n+{\rm i}\kappa) (n+{\rm i}\kappa)}},
\label{eqa10}
\end{eqnarray}
The other coefficients $r_2$ and $t_2$ at the foil-vacuum boundary can be obtained similarly. Next, we consider the wave vectors inside the membrane, $\boldsymbol{k}_{\rm{f1}}$ and $\boldsymbol{k}_{\rm{f2}}$. The magnitudes of their real and imaginary parts are given by
\begin{eqnarray}
&&|\boldsymbol{k}_{\rm{f1}}^{\rm{R}}|= |\boldsymbol{k}_{\rm{f2}}^{\rm{R}}| \nonumber \\
&&= \frac{\omega}{c}\sqrt{\sin^2 \theta_{\rm in} + {\rm Re} \Bigl \{ \sqrt{(n+{\rm i}\kappa)(n+{\rm i}\kappa)- \sin^2 \theta_{\rm in}} \Bigr \} }, \nonumber \\
&&
\label{eqa11}
\end{eqnarray}
\begin{eqnarray}
&&|\boldsymbol{k}_{\rm{f1}}^{\rm{I}}| =|\boldsymbol{k}_{\rm{f2}}^{\rm{I}}|  \nonumber \\
&&= \frac{\omega}{c} {\rm Im} \Bigl \{ \sqrt{(n+{\rm i}\kappa)(n+{\rm i}\kappa)- \sin^2 \theta_{\rm in}}  \Bigr  \},
\label{eqa12}
\end{eqnarray}
respectively. The imaginary part $\boldsymbol{k}_{\rm{f1}}^{\rm{I}}$ is perpendicular to foil surface and the real part $\boldsymbol{k}_{\rm{f1}}^{\rm{R}}$ points to the direction given by the angle of $\theta_{\rm{t}} = {\rm{asin}}(\omega \sin \theta_{\rm{in}}/c|\boldsymbol{k}_{\rm{f1}}^{\rm{R}}|)$ from the surface normal. By using the equations above, one obtains the electric fields as shown in Figs. \ref{fig7}(a)-(c). Magnetic fields are obtained by taking the real parts of the cross products between the complex forms of wave vectors and electric fields.

In order to calculate the momentum gains with Eqs. \ref{eq2}-\ref{eq4}, we need to perform time integrals of these electromagnetic fields. In addition to Eq. \ref{eqa4}, we can use two other types of integrals,
\begin{eqnarray}
&&\int_{t_a+t_0}^{t_b+t_0} e^{-\boldsymbol{k}^{\rm{I}} \boldsymbol{x}} \cos( \boldsymbol{k}^{\rm{R}} \boldsymbol{x} - \omega t + \phi) {\rm d}t \nonumber \\
&&= \Biggl[ \frac{e^{-k_z^{\rm I}v_0t}}{{(k_z^{\rm I}v_0)^2 + (\omega - k_z^{\rm R}v_0)}^2} \nonumber \\
&& \times \Bigl \{ -k_z^{\rm I}v_0 \cos(\omega t_0 +(\omega - k_z^{\rm R}v_0)t -\phi) \nonumber \\
&& + (\omega - k_z^{\rm R}v_0)\sin(\omega t_0 +(\omega - k_z^{\rm R}v_0)t -\phi ) \Bigr \} \Biggr]_{t_a}^{t_b},
\label{eqa13}
\end{eqnarray}
\begin{eqnarray}
&&\int_{t_a+t_0}^{t_b+t_0} e^{-\boldsymbol{k}^{\rm{I}} \boldsymbol{x}} \sin( \boldsymbol{k}^{\rm{R}} \boldsymbol{x} - \omega t + \phi) {\rm d}t \nonumber \\
&&= \Biggl[ \frac{e^{-k_z^{\rm I}v_0t}}{{(k_z^{\rm I}v_0)^2 + (\omega - k_z^{\rm R}v_0)}^2} \nonumber \\
&& \times \Bigl \{ k_z^{\rm I}v_0 \sin(\omega t_0 +(\omega - k_z^{\rm R}v_0)t -\phi) \nonumber \\
&& + (\omega - k_z^{\rm R}v_0)\cos(\omega t_0 +(\omega - k_z^{\rm R}v_0)t -\phi ) \Bigr \} \Biggr]_{t_a}^{t_b}.
\label{eqa14}
\end{eqnarray}
For strongly absorbing materials and thick membranes it is sufficient to consider only three waves, namely the incident and reflected fields on the laser input side and a decaying field inside the membrane. For metallic foils, one can further neglect the field inside of membranes.

\section{CLASSICAL ENERGY SPECTRUM AND ITS FOURIER TRANSFORMATION \label{seca4}}
\renewcommand{\theequation}{A\arabic{equation}}
\setcounter{equation}{14}
Here we estimate the shortest possible electron pulse duration by a hybrid quantum/classical approach. We consider an energy spectrum of classical point particles and then Fourier-transform a wave packet with this spectrum back to the time domain.

As discussed in the main text, the amount of the energy gain $\Delta W$ is proportional to the longitudinal momentum shift $\Delta p_z$ if $|\Delta p_z | \ll p_0$. Because $\Delta p_z$ sinusoidally oscillates with the laser-electron timing $t_0$ (see Appendix \ref{seca1}), the classical electron energy spectrum $I(\Delta W)$  is given by the probability density of a sinusoidal function\cite{blair1994}, 
\begin{eqnarray}
I(\Delta W) = c_W \frac{1}{\sqrt{1-(2\Delta W/\Delta W_{\rm max})^2}}
\label{eqa15}
\end{eqnarray}
where $c_W$ is a constant, $\Delta W$ is the amount of energy shift $(-\Delta W_{\rm{max}}/2 < \Delta W <  \Delta W_{\rm{max}}/2)$ and $\Delta W_{\rm{max}}/2$ is the maximum energy gain. We note that the energy spectrum in quantum mechanics shows pronounced sideband peaks at multiples of the photon energy\cite{feist2015,kirchner2014,barwick2009,ishida2017}, but these are neglected here. In order to achieve the probability density in the time domain, we consider a wavepacket with an energy spectrum as given by Eq. \ref{eqa15} and with a flat phase. With the help of the Mathematica software, we obtain the temporal structure $I(t)$ which is given by 
\begin{eqnarray}
I(t) = c_t \Big [ {}_0F_1(; \frac{5}{4}; -\Big( \frac{\Delta W_{\rm max} t} {4\hbar} \Bigr)^2   \Big ]
\label{eqa16}
\end{eqnarray}
where $c_t$ is a constant, $_0F_1$ is the hypergeometric function and $\hbar$ is Planck constant. Because $ \left[ _0F_1 (;5/4;-z^2) \right]^2$ has the half maximum at $|z|\sim0.63$, we obtain an energy-time relationship of
\begin{eqnarray}
\Delta W_{\rm max} \Delta t_{\rm{limit}} \approx 5 \hbar
\label{eqa17}
\end{eqnarray}
where $\Delta t_{\rm{limit}}$ is the FWHM width of temporal structure$I(t)$. Interestingly, this almost-classical result is close to $\Delta W_{\rm{max}} \Delta t_{\rm{limit}} \approx 4\hbar$, which was found with full quantum mechanical simulations\cite{baum2017}.

\section{VELOCITY MISMATCH AND ELECTRON PULSE TILT \label{seca5}}

\begin{figure}[!tb]
\includegraphics[width=8.5cm]{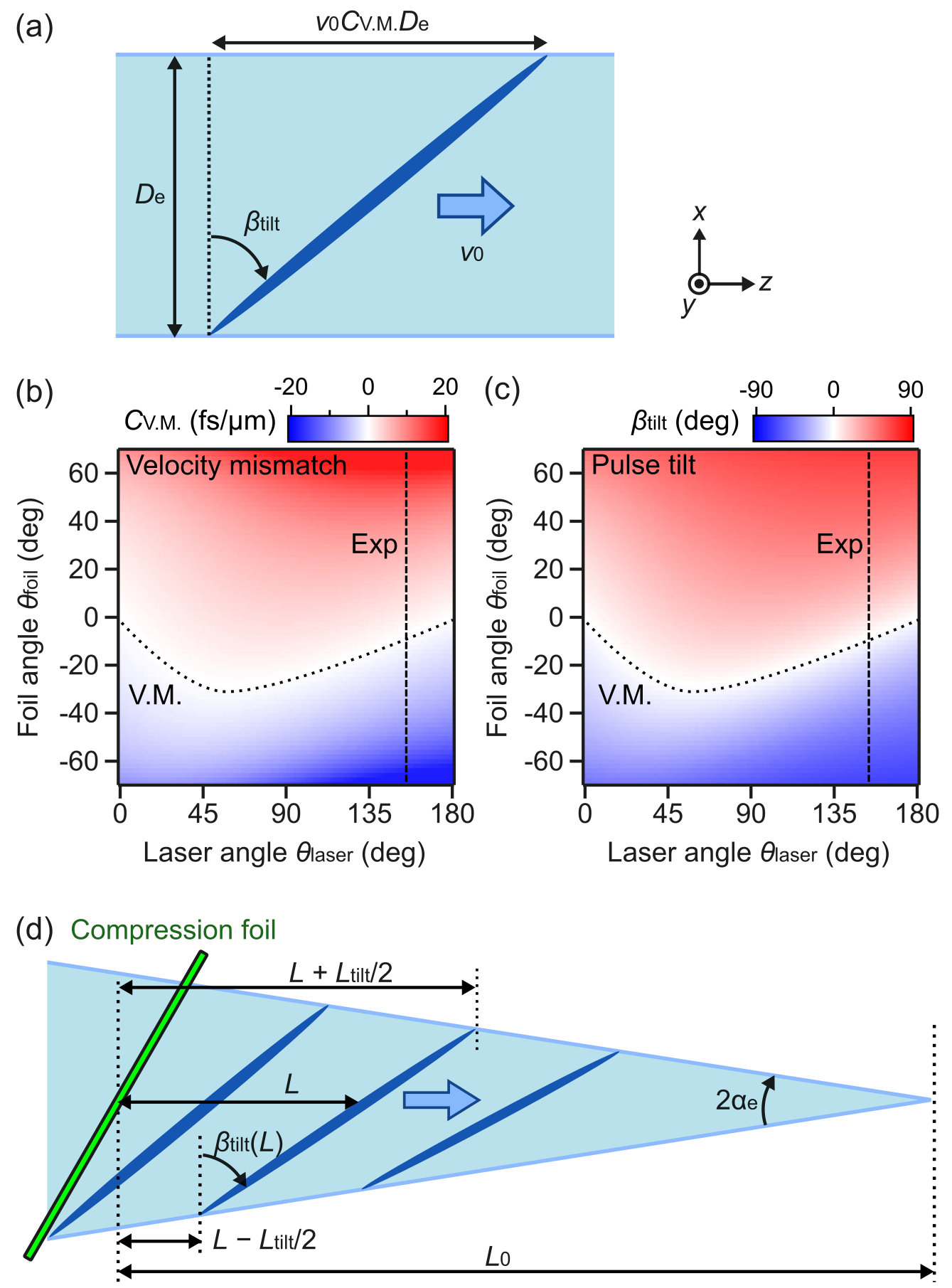}%
\caption{Velocity mismatch and pulse front tilt. (a) Definition of tilt of compressed electrons (see Appendix \ref{seca5}). (b) Velocity-mismatching effect as a function of laser and foil angles. (c) Tilt angle. Dotted line, velocity-matching condition; dashed line, condition in most of the reported experiments. (d) Tilt changes by converging/diverging beams. \label{fig10}}
\end{figure}

\renewcommand{\theequation}{A\arabic{equation}}
\setcounter{equation}{17}
As shown in Section \ref{sec4} and Fig. \ref{fig6}(c), the peak current density of the compressed electron pulses is tilted with respect to the propagation direction of the electron beam. This tilting in the experiment is due to the electron-laser velocity mismatch effect, i.e., the variation of laser-electron timing over a finite electron beam diameter\cite{williamson1993,baum2006}. We assumed a plane-wave laser field and an infinitely thin sample. The timing variation per unit electron beam size along $x$-axis is given by
\begin{eqnarray}
c_{\rm V.M.} = \frac{ \tan\theta_{\rm foil}}{v_0} + \frac{ \sin(\theta_{\rm laser} -\theta_{\rm foil})}{c \cos\theta_{\rm foil}}
\label{eqa18}
\end{eqnarray}
where $\theta_{\rm{foil}}$ and $\theta_{\rm{laser}}$ are foil and laser angles, see Fig. \ref{fig2}(a) for their definitions. The first and second terms on the right side of Eq. \ref{eqa18} represent the contributions of the electron and laser beams, respectively. From Eq. \ref{eqa18}, the condition for velocity-matching\cite{kirchner2014}, $c_{\rm{V.M.}}=0$, is given by
\begin{eqnarray}
\frac{v_0}{c} = \frac{ \sin\theta_{\rm foil} } { \sin(\theta_{\rm foil} -\theta_{\rm laser}) }
\label{eqa19}
\end{eqnarray}
At angle combinations satisfying this velocity matching, we can accelerate and compress electrons without any sideways deflections (see Section \ref{sec2}) and without pulse tilting. On the other hand, at other configurations, the electron pulses will be titled. For a collimated electron beam, the tilt angle $\beta_{\rm{tilt}}$ measured from $x$-axis (see Fig. \ref{fig10}(a)) is given by
\begin{eqnarray}
\beta_{\rm tilt} = {\rm atan} (c_{\rm V.M.} v_0)
\label{eqa20}
\end{eqnarray}
Figures \ref{fig10}(b) and \ref{fig10}(c) show the calculated velocity mismatch effect and the corresponding tilt angles, respectively, for 70-keV electrons. At our experimental attosecond compression configuration in Section \ref{sec4} with $\theta_{\rm{foil}} = 35^\circ$ and $\theta_{\rm{laser}}=155^\circ$, we expect a tilt of  $\beta_{\rm{tilt}} = 50^\circ$ . This tilt must be compensated by the characterization stage.

A residual divergence or convergence of the electron beam can reduce or enhance the tilt. In our experiment, the electron beam was slightly focused through both membranes onto the screen at a distance $L_0$. In order to estimate the degree of tilt enhancement, we consider an electron beam focused to $L_0$ with a half angle of $\alpha_{\rm{e}}$, see Fig. \ref{fig10}(d). The tilt angle $\beta_{\rm{tilt}}$ at any distance $L$ from the compression foil is given by
\begin{eqnarray}
\beta_{\rm tilt}(L) &&= \frac{\pi}{2} - \alpha_{\rm e} \nonumber \\
&& + {\rm asin} \Biggl ( \sqrt{ \frac{ (L_0 - L - L_{\rm tilt}/2)^2 \sin^2(2 \alpha_{\rm e})}{ 4\sin^2 \alpha_{\rm e} ( (L_0 - L)^2 - L_{\rm tilt}^2/2 ) + L_{\rm tilt}^2  }   } \Biggr ) \nonumber \\
&&
\label{eqa21}
\end{eqnarray}
where $L_{\rm{tilt}}$ is the path difference due to velocity mismatching, and expressed as $L_{\rm{tilt}} = v_0 c_{\rm{V.M.}} D_{\rm{e}}/\cos{\alpha_{\rm{e}}}$ with electron beam diameter $D_{\rm{e}}$. At our experimental conditions with $\alpha_{\rm{e}}= 0.05$ mrad, $L_0\approx 1.3$ m, $L_{\rm{tilt}}=150$ $\upmu$m and a distance between the compression and the characterization membranes of 4 mm, we obtain $\Delta \beta =\beta_{\rm{tilt}}(0 \ \rm{mm})-\beta_{\rm{tilt}}(4 \ \rm{mm})=0.09^\circ$. This slight change of the tilt blurs the electron's arrival time at the streaking foil by $\Delta \beta D_{\rm{e}}/v_0 \sim 1.4$ fs, and therefore needs to be compensated in the experiment for achieving attosecond resolution (see Section \ref{sec4} and Table \ref{table1}).

%



%


\end{document}